\documentclass{jpp}
\usepackage{graphicx}
\usepackage{epstopdf, epsfig}
\usepackage{hyperref}
\usepackage{amsmath}

\def\overar{\overset{\leftrightarrow}}

\shorttitle{Scattering of radio frequency waves in the plasma edge}
\shortauthor{S. I. Valvis, K. Hizanidis and P. Papagiannis}

\title{Scattering of radio frequency waves by cylindrical blobs in the plasma edge in tokamaks}

\author{S. I. Valvis\aff{1}
	\corresp{\email{jasonv@central.ntua.gr}},
	K. Hizanidis\aff{1}, P. Papagiannis\aff{1}, A. Papadopoulos\aff{1}, E. Glytsis\aff{1}, A. Zisis\aff{2}, I. G. Tigelis\aff{2}
	\and A. K. Ram\aff{3}}

\affiliation{\aff{1}School of Electrical and Computer Engineering, National Technical University of Athens, 9 Iroon Polytechniou Street, Athens 15780, GR
	\aff{2}Department of Physics, National and Kapodistrian University of Athens, University Campus, Zografou, Athens 15784, GR
	\aff{3}Plasma Science and Fusion Center, Massachusetts Institute of Technology, Cambridge MA, 175 Albany Street, Cambridge, MA 02139, USA}

\begin{document}
	
	\maketitle
	
	\begin{abstract}
		Radio frequency waves are routinely used in tokamak fusion plasmas for plasma heating, current control, and as well as in diagnostics. These waves  are excited by antenna structures placed near the tokamak's wall and they have to propagate through a turbulent layer known as scrape-off layer, before reaching the core plasma (which is their target). This layer exhibits coherent density fluctuations in the form of blobs and filaments.
		The scattering processes of RF plane waves by single blob is studied, with the assumption that the blob has cylindrical shape and infinite length. The axis of the blob is not necessarily aligned with the externally applied magnetic field in the case of plane waves.
		The investigation concerns the case of Electron Cyclotron (EC) waves ($f_0$=170 $GHz$) for ITER-like and Medium Size Tokamak applications (such as TCV, ASDEX-U, DIII-D, etc) as well as the case of Low Hybrid (LH) waves ($f_0$=4.6 $GHz$) for ALCATOR C-MOD Tokamak device. The study covers for a variety of density contrasts between the blobs and the ambient plasma and a wide range of blob radii.
		
	\end{abstract}
	
	\section{Introduction}
	An external antenna structure at the edge of a tokamak fusion device, excites radio frequency waves for plasma heating and plasma current generating. These waves, have to propagate through a turbulent scrape-off layer before they reach their target, which is the plasma core. The scrape-off layer consists of blobs and filaments surrounded by the ambient plasma, which differ from their background environment  with respect to the plasma electrons density. As a result, the plasma permittivity of this layer's filaments is different from the background plasma's permittivity. For this reason, the characteristic properties of the incident radio frequency waves can change during their transition through the scrape-off layer, so that the RF waves are modified by the fluctuations inside it.
	\par The study of the effects that the propagation through a different dielectric media (like blobs) has on the incident's wave properties (like electric field's intensity, Poynting vector's direction etc.) could be studied by using the geometric optics approximation. However, there is a limitation for the geometric optic's results to be valid: the electrons density of the filament $n_{fila}$ must be near equal to the ambient electrons density $n_{ambi}$, so that the relative density contrast between the blob and the ambient electrons density is $\delta n\equiv |n_{fila}-n_{ambi}|/n_{fila}\ll 1$, which does not happen. A typical experimental range of values for $\delta n$ is inside $(0.05,1)$. So, one can understand that there is a physical reason to derive a more general approach with validity in a larger domain which of course, must include the electron densities domain of interest.
	\par In this paper, Maxwell's equations and other mathematical tools (like Fourier transformation, Bessel functions, cylindrical vector functions, etc.) are used to derive a full-wave analytical model for the scattering process of RF waves by cylindrical density filaments. The study described, follows the study of the previous 
	\textit{Scattering of radio frequency waves by cylindrical density filaments in tokamak plasmas (Ram A. K., Hizanidis K.),} 
	but has more general validity: There is no limitation for the axis of the cylindrical filament to be aligned to the externally imposed magnetic field. The magnetic field is in angle $\phi_0$ with respect to the axis of the cylinder. It must be noted that for the purposes of this paper, the toroidal plasma is assumed to have a large aspect ratio, approximated by an infinitely extended slab where the filaments exist and the magnetic field is homogeneous. In addition, the axis of the cylindrical blob is assumed to have infinite length, so that the effects due to the ends of a filament of finite axial length can be ignored. Also, the thermal effects in the scrape-off layer are not taken into account, in order for the background ambient plasms as well as the filament to be assumed cold and uniform with the permittivity being the one of a cold plasma.
	\par Except of the larger ranges of validity for the electrons densities relative contrast between the interior and the exterior of the blob region that the full-wave model offers, there are in parallel some physical phenomena included, that the geometric optics approximation does not describe. First of all, except of refraction, in the full-wave model reflection, diffraction and shadowing are also studied. More, the fluctuations can couple power to other plasma waves (e.g. an ordinary mode wave can activate an extraordinary mode one and vice versa). In addition, the scattered waves propagate in all radial directions of the cylindrical filament and not only forward to the core, so that there are losses on the incident power.
	\par Another assumption of different kind that is made in this study, is that the fluctuations in the scrape-off layer are static, which means that they stand still where they are. In fact, the fluctuations are moving but there is nothing wrong with the assumption that they do not move, while according to experimental results, the speed of the toroidal propagation of the fluctuations around the tokamak is about $5*10^3$ $m/s$, about five orders of magnitude below the RF waves propagation speed which is nearly the well-known speed of light.
	\par Concerning the structure of this paper and the procedure of this study, by starting from a geometrical description, the Maxwell's equations, the dispersion relation and by making a mathematical analysis with the help of some mathematical tools (like Fourier transformations, coordinate systems transformations, cylindrical vector functions, etc), the Poynting vector in the forward (to the plasma core) direction is exported. In parallel, the electric and magnetic field components as well as the Poynting vector rest components are calculated. This analytical study, is followed by a numerical one, in which the way the fluctuations affect the RF waves propagation to the core is presented. The numerical study, covers a variety of different blob radii, different magnetic field inclinations with respect to the filament axis and different relative density contrasts between the blob and the ambient electrons density. It must be also noted that, the results received from an incident radio frequency wave of O-mode are different compared to the ones from the X-mode radio frequency wave.
	
	\section{Useful formulas for the geometry}
	\subsection{Transforming from the magnetic field coordinate system to the infinite-length cylindrical filament coordinate system}
	We first consider the magnetic field lines to be parallel to the $(z-x)$-plane with $z$ being the axis of the cylindrical blob. The cylinder has radius $a$ and as mentioned in the introduction, is assumed to have infinite length. This assumption is acceptable, since the radius $a$ is much smaller than it's length. As a result, there are not taking place effects due to the ends of a finite-size filament. In addition, the electrons density inside the cylinder as well as the electrons density outside is considered as homogeneous and the plasma is assumed to be cold. It must be also mentioned that in this full-wave model, there is no limitation for the ratio of these two different densities. 
	\par Let $\phi_{0}$ be the angle of the magnetic field lines (axis $z'$) with respect to $z$. In order to transform between the Cartesian coordinate systems of the blob and the magnetic field one has to rotate around the fixed axis $y$ with the help of the turning matrix
	
	\begin{align}
	\overar{\mathbf{R}}_{y}(\phi_0)\equiv
	\begin{pmatrix}
	\cos\phi_0 & 0 & -\sin\phi_0\\
	0 & 1 & 0\\
	\sin\phi_0 & 0 & \cos\phi_0\\
	\end{pmatrix}
	\end{align}
	\begin{figure}
		\centerline{\includegraphics[scale=0.4]{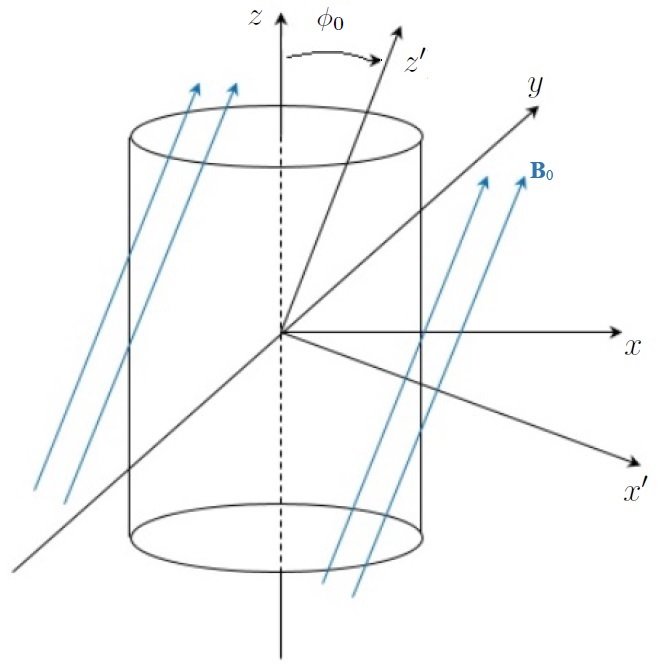}}
		\caption{Magnetic field and cylinder coordinate systems}
		\label{fig:ka}
	\end{figure}	
	to obtain:
	\begin{align*}
	\begin{pmatrix}
	\hat{\mathbf{x}}'\\
	\hat{\mathbf{y}}'\\
	\hat{\mathbf{z}}'\\
	\end{pmatrix}
	=\overar{\mathbf{R}}_{y}(\phi_{0})
	\begin{pmatrix}
	\hat{\mathbf{x}}\\
	\hat{\mathbf{y}}\\
	\hat{\mathbf{z}}\\
	\end{pmatrix},  
	\begin{pmatrix}
	\hat{\mathbf{x}}\\
	\hat{\mathbf{y}}\\
	\hat{\mathbf{z}}\\
	\end{pmatrix}
	=\overar{\mathbf{R}}_{y}(-\phi_{0})
	\begin{pmatrix}
	\hat{\mathbf{x}}'\\
	\hat{\mathbf{y}}'\\
	\hat{\mathbf{z}}'\\
	\end{pmatrix}\equiv
	\overar{\mathbf{R}}_{y}^{-1}(\phi_{0})
	\begin{pmatrix}
	\hat{\mathbf{x}}'\\
	\hat{\mathbf{y}}'\\
	\hat{\mathbf{z}}'\\
	\end{pmatrix}\
	\end{align*}
	where the primed unit vectors refer to the magnetic field line coordinate system with the same $y$-axis. Similarly, for any vector $\mathbf{a}$:
	\begin{align*}
	\begin{pmatrix}
	a_x'\\
	a_y'\\
	a_z'\\
	\end{pmatrix}
	=\overar{\mathbf{R}}_{y}(\phi_{0})
	\begin{pmatrix}
	a_x\\
	a_y\\
	a_z\\
	\end{pmatrix},  
	\begin{pmatrix}
	a_x\\
	a_y\\
	a_z\\
	\end{pmatrix}
	=\overar{\mathbf{R}}_{y}(-\phi_{0})
	\begin{pmatrix}
	a_x'\\
	a_y'\\
	a_z'\\
	\end{pmatrix}\equiv
	\overar{\mathbf{R}}_{y}^{-1}(\phi_{0})
	\begin{pmatrix}
	a_x'\\
	a_y'\\
	a_z'\\
	\end{pmatrix}\
	\end{align*}
	and
	\begin{align*}
	\begin{pmatrix}
	a_x' & a_y' & a_z'
	\end{pmatrix}
	=\begin{pmatrix}
	a_x & a_y & a_z
	\end{pmatrix}
	\overar{\mathbf{R}}_{y}(-\phi_{0})\equiv
	\begin{pmatrix}
	a_x & a_y & a_z
	\end{pmatrix}
	\overar{\mathbf{R}}_{y}^{-1}(\phi_{0})
	\end{align*}  
	\begin{align*}
	\begin{pmatrix}
	a_x & a_y & a_z
	\end{pmatrix}
	=\begin{pmatrix}
	a_x' & a_y' & a_z'
	\end{pmatrix}
	\overar{\mathbf{R}}_{y}(\phi_{0})
	\end{align*}
	\subsection{Transforming from Cartesian to cylindrical coordinates and vice-versa}
	While being in the cylinder (blob) - based frame of reference, one can transform from cartesian coordinate system to cylindrical coordinates in the same frame. Thus, by using the transformation matrix
	\begin{align*}
	\overar{\mathbf{R}}_c(\varphi_k)=
	\begin{pmatrix}
	\cos\varphi_k & -\sin\varphi_k & 0\\
	\sin\varphi_k & \cos\varphi_k & 0\\
	0 & 0 & 1
	\end{pmatrix}
	\end{align*}
	one obtains:
	\begin{align*}
	\begin{pmatrix}
	\hat{\mathbf{x}}\\
	\hat{\mathbf{y}}\\
	\hat{\mathbf{z}}\\
	\end{pmatrix}
	=\overar{\mathbf{R}}_c(\varphi_k)
	\begin{pmatrix}
	\hat{\mathbf{r}}\\
	\hat{\boldsymbol{\varphi}}\\
	\hat{\mathbf{z}}\\
	\end{pmatrix},  
	\begin{pmatrix}
	\hat{\mathbf{r}}\\
	\hat{\boldsymbol{\varphi}}\\
	\hat{\mathbf{z}}\\
	\end{pmatrix}
	=\overar{\mathbf{R}}_c(-\varphi_k)
	\begin{pmatrix}
	\hat{\mathbf{x}}\\
	\hat{\mathbf{y}}\\
	\hat{\mathbf{z}}\\
	\end{pmatrix}\equiv
	\overar{\mathbf{R}}_c^{-1}(\varphi_k)
	\begin{pmatrix}
	\hat{\mathbf{x}}\\
	\hat{\mathbf{y}}\\
	\hat{\mathbf{z}}\\
	\end{pmatrix}\
	\end{align*}
	Here, $\varphi_k$ refers to the azimuthal angle in the blob-based coordinate system. Similarly, for any vector $\mathbf{a}$:
	\begin{align*}
	\begin{pmatrix}
	a_x\\
	a_y\\
	a_z\\
	\end{pmatrix}
	=\overar{\mathbf{R}}_c(\varphi_k)
	\begin{pmatrix}
	a_r\\
	a_\varphi\\
	a_z\\
	\end{pmatrix},  
	\begin{pmatrix}
	a_r\\
	a_\varphi\\
	a_z\\
	\end{pmatrix}
	=\overar{\mathbf{R}}_c(-\varphi_k)
	\begin{pmatrix}
	a_x\\
	a_y\\
	a_z\\
	\end{pmatrix}\equiv
	\overar{\mathbf{R}}_c^{-1}(\varphi_k)
	\begin{pmatrix}
	a_x\\
	a_y\\
	a_z\\
	\end{pmatrix}\
	\end{align*}
	and
	\begin{align*}
	\begin{pmatrix}
	a_x & a_y & a_z
	\end{pmatrix}
	=\begin{pmatrix}
	a_r & a_\varphi & a_z
	\end{pmatrix}
	\overar{\mathbf{R}}_c(-\varphi_k)\equiv
	\begin{pmatrix}
	a_r & a_\varphi & a_z
	\end{pmatrix}
	\overar{\mathbf{R}}_c^{-1}(\varphi_k)
	\end{align*}  
	\begin{align*}
	\begin{pmatrix}
	a_r & a_\varphi & a_z
	\end{pmatrix}
	=\begin{pmatrix}
	a_x & a_y & a_z
	\end{pmatrix}
	\overar{\mathbf{R}}_c(\varphi_k)
	\end{align*}

	\section{Incident wave}
	The wavevector of the RF wave is symbolized $\mathbf{k}$ and the radial position vector $\mathbf{r}$. By normalizing these two vectors to the dimensionless ones $\boldsymbol{\eta} \equiv \mathbf{k}c/\omega$ and $\boldsymbol{\rho} \equiv \mathbf{r}\omega/c$ with $c$ being the speed of light in vacuum and  $\omega=2 \pi f$ being the angular frequency (while $f$ is the frequency). The dot product of $\boldsymbol{\eta}\cdot\boldsymbol{\rho}$ (which is the normalized $\mathbf{k}\cdot\mathbf{r}$) can be calculated in cylindrical coordinates as:
	\begin{eqnarray*}
		\boldsymbol{\eta}\cdot\boldsymbol{\rho}&=&\begin{pmatrix}
			\chi & \psi & \zeta
		\end{pmatrix}\begin{pmatrix}
			\eta_x\\ \eta_y\\ \eta_z
		\end{pmatrix}\\
		&=& \begin{pmatrix}
			\rho\cos\varphi & \rho\sin\varphi & \zeta
		\end{pmatrix}\overar{\mathbf{R}}_c(\varphi_k)\begin{pmatrix}
			\eta_r\\ 0\\ \eta_z
		\end{pmatrix} =
		\eta_r\rho\cos(\varphi-\varphi_k)+\eta_{\zeta}\zeta
	\end{eqnarray*}
	where $\varphi$ and $\varphi_k$ are the azimuthal angles between the $x$-axis and $\boldsymbol{\rho}$ and $\boldsymbol{\eta}$, respectively.
	From now on, the subscript index "$0$" is used when referring to the incident wave, for which one may write:
	\begin{align*}
	\boldsymbol{\eta}_0\cdot\boldsymbol{\rho}=\eta_{0r}\rho\cos(\varphi-\varphi_{0k})+\eta_{0z}\zeta
	\end{align*}
	where, now, the azimuthal angle is not in general zero. The normalized to its amplitude incident electric field intensity is:
	\begin{align*}
	\frac{\mathbf{E}_0}{E_0}\exp(i\boldsymbol{\eta}_0\cdot\boldsymbol{\rho})\equiv\mathbf{e}_0^P\exp\left\{i\left[\eta_{0r}\rho\cos(\varphi-\varphi_{0k}+\eta_{0z}\zeta\right]\right\}
	\end{align*}
	In terms of the exponential dyadic involving the vector cylinder functions $\mathbf{m}_n, \quad \mathbf{n}_n, \quad \mathbf{l}_n$ in the cylinder frame of reference (convenient) one has
	\begin{align*}
	\begin{split}
	\mathbf{e}_0^P\exp(i\boldsymbol{\eta}_{0r}\cdot\boldsymbol{\rho})=\mathbf{e}_0^P\cdot\sum_{n=-\infty}^{n=\infty}&\left[\mathbf{a}_{0n}\mathbf{m}_n(\eta_{0r}\rho,\eta_{0z}\zeta,\varphi)+\mathbf{b}_{0n}\mathbf{n}_n(\eta_{0r}\rho,\eta_{0z}\zeta,\varphi)\right.\\
	&\left.+\mathbf{c}_{0n}\mathbf{l}_n(\eta_{0r}\rho,\eta_{0z}\zeta,\varphi)\right]
	\end{split}
	\end{align*}
	The cylindrical vector functions in cylindrical coordinates are as follows:
	\begin{align*}
	\mathbf{m}_n(\eta_{r}\rho,\eta_{z}\zeta,\varphi)\equiv\left[in\frac{Z_n(\eta_r\rho)}{\rho}\hat{\mathbf{r}}-\frac{dZ_n(\eta_r\rho)}{d\rho}\hat{\boldsymbol{\varphi}}\right]\exp\left[i(\eta_z\zeta+n\varphi)\right]
	\end{align*}
	\begin{align*}
	\mathbf{n}_n(\eta_{r}\rho,\eta_{z}\zeta,\varphi)\equiv\left\{\frac{\eta_z}{\eta}\left[i\frac{dZ_n(\eta_r\rho)}{d\rho}\hat{\mathbf{r}}-n\frac{Z_n(\eta_r\rho)}{\rho}\hat{\boldsymbol{\varphi}}\right]+\frac{\eta_r^2}{\eta}Z_n(\eta_r\rho)\hat{\boldsymbol{\zeta}}\right\}\exp\left[i(\eta_z\zeta+n\varphi)\right]
	\end{align*}
	and
	\begin{align*}
	\mathbf{l}_n(\eta_{r}\rho,\eta_{z}\zeta,\varphi)\equiv\left[\frac{dZ_n(\eta_r\rho)}{d\rho}\hat{\mathbf{r}}+in\frac{Z_n(\eta_r\rho)}{\rho}\hat{\boldsymbol{\varphi}}+i\eta_zZ_n(\eta_r\rho)\hat{\boldsymbol{\zeta}}\right]\exp\left[i(\eta_z\zeta+n\varphi)\right]
	\end{align*}
	
	Note that these vector functions are expressed in terms of the position in space (in cylindrical coordinates) while the wave enters only via its axial and radial refractive index in the cylinder frame of reference. For the incident wave the Bessel functions involved are $J_n$. The vector functions obey the following relations:
	\begin{align*}
	\nabla\cdot\mathbf{m}_n=0, \quad \nabla\cdot\mathbf{n}_n=0, \quad \nabla\cdot\mathbf{l}_n=-\eta^2Z_n\exp\left[i(\eta_z\zeta+n\varphi)\right]
	\end{align*}
	as well as:
	\begin{align*}
	\nabla\times\mathbf{l}_n=0, \quad \nabla\times\mathbf{m}_n=\eta\mathbf{n}_n, \quad \nabla\times\mathbf{n}_n=\eta\mathbf{m}_n
	\end{align*}
	\subsection{Calculations of the vectors $\mathbf{a}_n$, $\mathbf{b}_n$, $\mathbf{c}_n$ in general}
	Because of the completeness property of the vector cylindrical functions, the vectorial coefficients $\mathbf{a}_n$, $\mathbf{b}_n$ and $\mathbf{c}_n$ in the dyadic of the exponential can be calculated for the incident or the blob fields and the applied as well in the expressions for the scattered fields. Note that the index "\textit{k}" is referring to the cylindrical filament's coordinate system. So, in the Cartesian coordinate system of the cylinder it is:
	\begin{align*}
	\begin{pmatrix}
	\hat{\mathbf{x}}\\ \hat{\mathbf{y}}\\ \hat{\mathbf{z}}
	\end{pmatrix}\text{e}^{i\boldsymbol{\rho}\cdot\boldsymbol{\eta}_k}=\sum_{n=-\infty}^{n=\infty}\left[\begin{pmatrix}
	a_{kn}^x\\ a_{kn}^y\\ a_{kn}^z
	\end{pmatrix}\mathbf{m}_n+\begin{pmatrix}
	b_{kn}^x\\ b_{kn}^y\\ b_{kn}^z
	\end{pmatrix}\mathbf{n}_n+\begin{pmatrix}
	c_{kn}^x\\ c_{kn}^y\\ c_{kn}^z
	\end{pmatrix}\mathbf{l}_n\right]
	\end{align*}
	
	where after calculations, the coefficients are: 
	\begin{align*}
		\begin{pmatrix}
			a_{kn}^x\\ a_{kn}^y\\ a_{kn}^z
		\end{pmatrix}=\begin{pmatrix}
			-\sin\varphi_k\\ \cos\varphi_k\\ 0
		\end{pmatrix}\dfrac{i^{n+1}\text{e}^{-in\varphi_k}}{\eta_{kr}}
	\end{align*}

	\begin{align*}
		\begin{pmatrix}
			b_{kn}^x\\ b_{kn}^y\\ b_{kn}^z
		\end{pmatrix}=-i^n\dfrac{\eta_{kz}}{\eta_k\eta_{kr}}\text{e}^{-in\varphi_k}\begin{pmatrix}
			\cos\varphi_k\\ \sin\varphi_k\\ -\frac{\eta_{kr}}{\eta_{kz}}
	\end{pmatrix}
	\end{align*}

	\begin{align*}
	\begin{pmatrix}
	c_{kn}^x\\ c_{kn}^y\\ c_{kn}^z
	\end{pmatrix}=-i^{n+1}\dfrac{\text{e}^{-in\varphi_k}}{\eta_k^2}\begin{pmatrix}
	\eta_{kr}\cos\varphi_k\\ \eta_{kr}\sin\varphi_k\\ \eta_{kz}
	\end{pmatrix}
	\end{align*}

	\section{Propagation of RF waves in plasma and the dispersion relation}
	\subsection{The electric field in general}
	For a cold plasma, the Faraday equation can be combined with the Ampere equation  in the Fourier domain to produce the following:
	\begin{align*}
	\varepsilon_0\nabla\times\nabla\times\mathbf{E(r)}-\left(\dfrac{\omega}{c}\right)^2\mathbf{D(r)}=0
	\end{align*}
	It is assumed that the plasma equilibrium is time independent and the linearized perturbed electromagnetic fields have a time dependence of the form $\text{e}^{-i \omega t}$, with $t$ being the time.
	In normalized wave vector representation:
	\begin{align*}
	\mathbf{E(r)}=\int\int\int d^3\eta\mathbf{E(\boldsymbol{\eta})}\exp(i\boldsymbol{\eta\cdot\rho})
	\end{align*}
	The combination of these two equations leads to:
	\begin{align*}
	\mathbf{E(r)}=\int\int\int d^3\eta\overar{\boldsymbol{\Delta}}(\boldsymbol{\eta})\mathbf{E(\boldsymbol{\eta})}\exp(i\boldsymbol{\eta\cdot\rho})
	\end{align*}
	where the dispersion tensor $\overar{\boldsymbol{\Delta}}(\boldsymbol{\eta})$ appears. For non-trivial solutions for the electric field intensity, the determinant of the dispersion tensor must be zero. The latter requirement, the dispersion relation in other words, selects the sub-manifold in the Fourier space where non-trivial electric field Fourier components exist. That is:
	\begin{align*}
	\boldsymbol{\mathcal{D}}(\boldsymbol{\eta})\equiv\det\left[\overar{\boldsymbol{\Delta}}(\boldsymbol{\eta})\right]=0
	\end{align*}
	or, in a cylindrical frame of reference for the wave vector with a $z$-axis aligned with the cylinder's $z$-axis:
	\begin{align*}
	\boldsymbol{\mathcal{D}}(\eta_{kr}, \eta_{kz}, \varphi_k)=0
	\end{align*}
	which it turns out to be a fourth order equation with respect to $\eta_{kr}$ (see in the following). Thus:
	\begin{align*}
	\mathbf{E}(\boldsymbol{\rho})=\sum_{M=1}^{4}\int_0^{2\pi}d\varphi_k\int_{-\infty}^{\infty}d\eta_{kz}\mathbf{E}_k\left[\eta_{kr}(\eta_{kz},\varphi_k),\eta_{kz},\varphi_k\right]\exp(i\boldsymbol{\eta_k\cdot\rho})
	\end{align*}
	or, equivalently:
	\begin{align*}
	\begin{split}
	\mathbf{E}(\rho, \varphi, \zeta)=&\sum_{M=1}^{4}\int_0^{2\pi}d\varphi_k\int_{-\infty}^{\infty}d\eta_{kz}\mathbf{E}_k^M\left[\eta_{kr}^M(\eta_{kz},\varphi_k),\eta_{kz},\varphi_k\right]\\
	&\exp\left\{i\left[\rho\eta_{kr}^M\cos(\varphi-\varphi_k)+\eta_{kz}\zeta\right]\right\}
	\end{split}
	\end{align*}
	with the letter "\textit{M}" referring to the number of the solution (while in general, there are four solutions for any fourth-order equation).
	One may introduce a factor of $\frac{1}{2\pi}$ in front of these expressions in order to realate easily back to the case of the aligned cylinder.
	
	\subsection{The dispersion relation exportation}
	In the magnetic field frame of reference (previously primed), in cartesian coordinates the permittivity tensor has the form:
	\begin{align*}
	\overar{\mathbf{K}}^{cart}_{mag}=
	\begin{pmatrix}
	K_{\perp} & -iK_{\times} & 0\\
	iK_{\times} & K_{\perp} & 0\\
	0 & 0 & K_{\parallel}
	\end{pmatrix}
	\end{align*}
	With the help of some formulas mentioned previously, the permittivity tensor can be expressed in the cylider frame of reference, in cartesian coordinates:
	\begin{align*}
	\overar{\mathbf{K}}^{cart}_{fila}=
	\overar{\mathbf{R}}^{-1}_y(\phi_0)
	\overar{\mathbf{K}}^{cart}_{mag}
	\overar{\mathbf{R}}_y(\phi_0)
	\end{align*}
	which after the calculations give:
	\begin{align*}
	\overar{\mathbf{K}}^{cart}_{fila}=
	\begin{pmatrix}
	K_{\perp}c^2_0+K_{\parallel}s^2_0 & -iK_{\times}c_0 & c_0s_0(K_{\parallel}-K_{\perp})\\
	iK_{\times}c_0 & K_{\perp} & -iK_{\times}s_0\\
	c_0s_0(K_{\parallel}-K_{\perp}) & iK_{\times}s_0 & K_{\perp}s^2_0+K_{\parallel}c^2_0\\
	\end{pmatrix}
	\end{align*}
	where
	\begin{align*}
	c_0\equiv\cos\phi_0, \quad  s_0\equiv\sin\phi_0
	\end{align*}
	It has to be emphasized, that the angle $\phi_0$ is the angle between the axis of the cylindrical filament and the magnetic field line and must not be confused with the azimuthal angles in the cylindrical coordinate systems, which in general are symbolized as $\varphi$. The dispersion tensor, can be calculated from the permittivity tensor, in the same frame of reference and the same coordinate system, by using the following:
	\begin{align*}
	\overar{\boldsymbol{\Delta}}^{cart}_{fila}=
	\overar{\mathbf{K}}^{cart}_{fila}+\big(\boldsymbol{\eta}\boldsymbol{\eta}-\mathbf{I}\eta^2\big)^{cart}_{fila}
	\end{align*}
	So, it is obtained that:
	\begin{align*}
	\overar{\boldsymbol{\Delta}}^{cart}_{fila}=
	\begin{pmatrix}
	K_{\perp}c^2_0+K_{\parallel}s^2_0-\eta^2_y-\eta^2_z & -iK_{\times}c_0+\eta_x\eta_y & c_0s_0(K_{\parallel}-K_{\perp})+\eta_x\eta_z\\
	iK_{\times}c_0+\eta_x\eta_y & K_{\perp}-\eta^2_x-\eta^2_z & -iK_{\times}s_0+\eta_y\eta_z\\
	c_0s_0(K_{\parallel}-K_{\perp})+\eta_x\eta_z & iK_{\times}s_0+\eta_y\eta_z & K_{\perp}s^2_0+K_{\parallel}c^2_0-\eta^2_x-\eta^2_y\\
	\end{pmatrix}
	\end{align*}
	Now, in cylindrical coordinates:
	
	\begin{eqnarray*}
		\overar{\boldsymbol{\Delta}}^{cyl}_{fila} &=&
		\overar{\mathbf{R}}^{-1}_{c}(\varphi_k)\overar{\boldsymbol{\Delta}}^{cart}_{fila}\overar{\mathbf{R}}_{c}(\varphi_k) \\ &=&
		\overar{\mathbf{R}}^{-1}_{c}(\varphi_k)\overar{\boldsymbol{K}}^{cart}_{fila}\overar{\mathbf{R}}_{c}(\varphi_k)
		+\boldsymbol{\eta}^{cyl}_{fila}\big(\boldsymbol{\eta}^{cyl}_{fila}\big)^T-\mathbf{I}\eta^2
	\end{eqnarray*}
	Note now that the index in the azimuthal angle refers to a particular wave vector in a $\mathbf{k}$-space coordinate system with $k_z$ component along the axis of the blob and azimuthal angle of the projection of $\mathbf{k}$ on the $(x-y)$ plane with respect to the $x$-axis of the Cartesian blob-based system. In the following the azimuthal component of the $\mathbf{k}$-field is actually zero. Executing the multiplications renders:
	\begin{align*}
	\big(\Delta^{cyl}_{fila}\big)_{11}=
	c_k^2\big(K_{\perp}c_0^2+K_{\parallel}s_0^2\big)+K_{\perp}s_k^2-\eta_{\varphi}^{2}-\eta_{z}^{2}
	\end{align*}
	\begin{align*}
	\big(\Delta^{cyl}_{fila}\big)_{12}=
	-s_kc_k\big(K_{\perp}c_0^2+K_{\parallel}s_0^2\big)+K_{\perp}s_kc_k-iK_{\times}c_0+\eta_r\eta_{\varphi}
	\end{align*}
	\begin{align*}
	\big(\Delta^{cyl}_{fila}\big)_{13}=
	c_kc_0s_0\big(K_{\parallel}-K_{\perp}\big)-iK_{\times}s_0s_k+\eta_r\eta_z
	\end{align*}
	\begin{align*}
	\big(\Delta^{cyl}_{fila}\big)_{21}=
	-s_kc_k\big(K_{\perp}c_0^2+K_{\parallel}s_0^2\big)+K_{\perp}s_kc_k+iK_{\times}c_0+\eta_r\eta_{\varphi}
	\end{align*}
	\begin{align*}
	\big(\Delta^{cyl}_{fila}\big)_{22}=
	s_k^2\big(K_{\perp}c_0^2+K_{\parallel}s_0^2\big)+K_{\perp}c_k^2-\eta_r^2-\eta_z^2
	\end{align*}
	\begin{align*}
	\big(\Delta^{cyl}_{fila}\big)_{23}=
	-s_kc_0s_0\big(K_{\parallel}-K_{\perp}\big)-iK_{\times}s_0c_k+\eta_{\varphi}\eta_z
	\end{align*}
	\begin{align*}
	\big(\Delta^{cyl}_{fila}\big)_{31}=
	c_kc_0s_0\big(K_{\parallel}-K_{\perp}\big)+iK_{\times}s_0s_k+\eta_r\eta_z
	\end{align*}
	\begin{align*}
	\big(\Delta^{cyl}_{fila}\big)_{32}=
	-s_kc_0s_0\big(K_{\parallel}-K_{\perp}\big)+iK_{\times}s_0c_k+\eta_{\varphi}\eta_z
	\end{align*}
	\begin{align*}
	\big(\Delta^{cyl}_{fila}\big)_{33}=
	\big(K_{\perp}s_0^2+K_{\parallel}c_0^2\big)-\eta_{r}^{2}-\eta_{\varphi}^{2}
	\end{align*}
	where:
	\begin{align*}
	c_k\equiv\cos\varphi_k, \quad  s_k\equiv\sin\varphi_k
	\end{align*}
	and the electric field cylindrical components satisfy the following
	\begin{align*}
	\overar{\boldsymbol{\Delta}}^{cyl}_{fila}
	\begin{pmatrix}
	E_r\\E_\varphi\\E_z
	\end{pmatrix}^{cyl}_{fila}=0
	\end{align*}
	As it was mentioned before, the azimuthal component for the propagation vector field $\mathbf{k}$ (the azimuthal angle $\varphi_k$ suffices) is zero. Thus, the dispersion relation written down previously in the blob cylindrical frame of reference can be simplified as follows:
	
	\begin{align*}
	\det\Big(\overar{\boldsymbol{\Delta}}^{cyl}_{fila}\Big)=0
	\end{align*}
	This, can be written also as:
	\begin{align*}
	\det\begin{pmatrix} Ac_k^2+K_{\perp}s_k^2-\eta_z^2 & (K_{\perp}-A)s_kc_k-iK_{\times}c_0 & Dc_k+\eta_r\eta_z-iK_{\times}s_0s_k\\
	(K_{\perp}-A)s_kc_k+iK_{\times}c_0 & As_k^2+K_{\perp}c_k^2-\eta_r^2-\eta_z^2 & -Ds_k-iK_{\times}s_0c_k\\
	Dc_k+\eta_r\eta_z+iK_{\times}s_0s_k & -Ds_k+iK_{\times}s_0c_k & A'-\eta_r^2
	\end{pmatrix}=0
	\end{align*}
	with:
	\begin{align*}
	\begin{split}
	&A\equiv K_{\perp}c_0^2+K_{\parallel}s_0^2, \quad A' \equiv K_{\perp}s_0^2+K_{\parallel}c_0^2, \quad D\equiv (K_{\parallel}-K_{\perp})s_0c_0, \quad s_k\equiv \sin\varphi_k,\\
	&c_k\equiv \cos\varphi_k
	\end{split}
	\end{align*}
	which, for the incident wave [the incident wave is assumed to have the propagation vector on the $(x-z)$-plane] becomes $(s_k=0, \quad c_k=1)$:
	\begin{align*}
	\det\begin{pmatrix} A-\eta_z^2 & -iK_{\times}c_0 & D+\eta_r\eta_z\\
	iK_{\times}c_0 & K_{\perp}-\eta_r^2-\eta_z^2 & -iK_{\times}s_0\\
	D+\eta_r\eta_z & iK_{\times}s_0 & A'-\eta_r^2
	\end{pmatrix}=0
	\end{align*}
	Note that for any mode $\eta_{kz}$ is preserved, that is, it is set by the incident one, $\eta_{0z}$. Therefore, one obtains, dropping the index $"k"$:
	\begin{align*}
	&\left[K_{\perp}-(K_{\perp}-K_{\parallel})s_0^2c^2\right]\eta_r^4\\
	&-2\eta_{0z}(K_{\perp}-K_{\parallel})s_0c_0c\eta_r^3\\
	&+\left\{(K_{\perp}+K_{\parallel})(\eta_{0z}^2-K_{\perp})+K_{\times}^2+\left[(K_{\perp}^2-K_{\times}^2-K_{\perp}K_{\parallel})c^2+\eta_{0z}^2(K_{\perp}-K_{\parallel})s^2\right]s_0^2\right\}\eta_r^2\\
	&-2\eta_{0z}\left[(K_{\perp}-K_{\parallel})(\eta_{0z}^2-K_{\perp})+K_{\times}^2\right]s_0c_0c\eta_r\\
	&+K_{\parallel}\left[(\eta_{0z}^2-K_{\perp})^2-K_{\times}^2\right]+\eta_{0z}^2\left[(K_{\perp}-K_{\parallel})(\eta_{0z}^2-K_{\perp})+K_{\times}^2\right]s_0^2=0
	\end{align*}
	For the incident wave in the inclined case and the propagation vector on the $z-z'$ plane (same as $x-z$ plane), that is for $c=1$, $s=0$, one obtains
	\begin{align*}
	&(K_{\perp}c_0^2+K_{\parallel}s_0^2)\eta_r^4\\
	&-2\eta_{0z}(K_{\perp}-K_{\parallel})s_0c_0\eta_r^3\\
	&+\left\{(K_{\perp}+K_{\parallel})(\eta_{0z}^2-K_{\perp})+K_{\times}^2+(K_{\perp}^2-K_{\times}^2-K_{\perp}K_{\parallel})s_0^2\right\}\eta_r^2\\
	&-2\eta_{0z}\left[(K_{\perp}-K_{\parallel})(\eta_{0z}^2-K_{\perp})+K_{\times}^2\right]s_0c_0\eta_r\\
	&+K_{\parallel}\left[(\eta_{0z}^2-K_{\perp})^2-K_{\times}^2\right]+\eta_{0z}^2\left[(K_{\perp}-K_{\parallel})(\eta_{0z}^2-K_{\perp})+K_{\times}^2\right]s_0^2=0
	\end{align*}
	For the aligned cylinder and when the propagation vector is on $x-z$ plane $(s_0=0, \quad c_0=1)$, the corresponding equation is:
	\begin{align*}
	K_{\perp}\eta_r^4+\left[(K_{\perp}+K_{\parallel})(\eta_{0z}^2-K_{\perp})+K_{\times}^2\right]\eta_r^2+K_{\parallel}\left[(\eta_{0z}^2-K_{\perp})^2-K_{\times}^2\right]=0
	\end{align*}
	or, equivalently, introducing the polar angle, $\vartheta$, representation:
	\begin{align*}
	\begin{split}
	&(K_{\perp}\sin^2\vartheta+K_{\parallel}\cos^2\vartheta)\eta_0^4-\left[(K_{\perp}^2-K_{\times}^2)\sin^2\vartheta+(1+\cos^2\vartheta)K_{\perp}K_{\parallel}\right]\eta_0^2\\
	&+(K_{\perp}^2-K_{\times}^2)K_{\parallel}=0
	\end{split}
	\end{align*}
	which is the standard well-known form (Stix). Back in the general case, we observe that, for a fixed $\eta_{0z}$ (which is preserved for an infinite cylinder along the $z$-axis) the equation posses four distinct roots for a particular choice of the angle $\varphi_k$, labeled $\eta_r^{(M)}$ with $M=1,2,3,4$. However, because of the presence of the cosine of the azimuthal angle in the odd order coefficients, these roots, viewed as functions of the azimuthal angle $\varphi_k$, are \textit{symmetric} with respect to the midpoint $\varphi_k=\pi$ and \textit{hetero-anti-symmetric} with respect to $\varphi_k=\frac{\pi}{2}$ and $\varphi_k=\frac{3\pi}{2}$. Thus, from one root function of $\varphi_k$, one may construct a second one by applying the aforementioned symmetries. We may name this pair as \textit{"symmetry-based pair"}. Since there exist four roots, there must exist two symmetry-based pairs. It is much more convenient to re-label the two pairs according to which one contains a member which coincides with the radial index of the incident wave (in the cylinder reference system). That is, one pair for the ambient environment will be labeled as O-pair (X-pair) if it contains the $\eta_{0r}$ of an incident O-mode (X-mode). The respective pair for the blob parameters will retain the same characterization in order to ensure that they coincide in the limit of zero contrast (between inside and outside). The remaining pair automatically will be labeled as X-pair (O-pair). Therefore, one may introduce the indices $O_1, O_2, X_1, X_2$ where now $O_1, O_2$ and $X_1, X_2$ being the two symmetry-based pairs. The symmetries are as follows $M=O,X$:
	
	\begin{figure}
		\centerline{\includegraphics[scale=1.0]{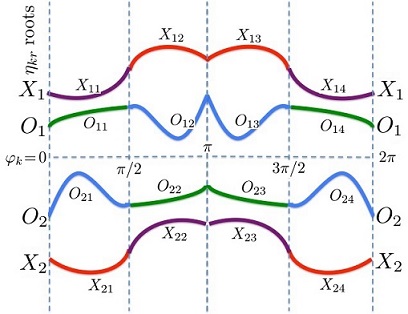}}
		\caption{Symmetry-based pairs (O and X)}
		\label{fig:ka}
	\end{figure}
	
	\begin{align*}
	\eta_{kr}^{M_i}(\varphi_k=0\to\pi)=-\eta_{kr}^{M_{3-i}}(\varphi_k=\pi\to0)
	\end{align*}
	and
	\begin{align*}
	\eta_{kr}^{M_i}(\varphi_k=0\to\pi)=-\eta_{kr}^{M_i}(\varphi_k=2\pi\to\pi)
	\end{align*}
	or, equivalently:
	\begin{align*}
	\eta_{kr}^{M_i}(\varphi_k)=\eta_{kr}^{M_i}(2\pi-\varphi_k)=-\eta_{kr}^{M_{3-i}}(\pi-\varphi_k)=-\eta_{kr}^{M_{3-i}}(\pi+\varphi_k)
	\end{align*}
	Of course, for $\varphi_k=\pi$, $\frac{3\pi}{2}$ the members of each pair are opposite (in the general sense).
	In the following, only one member of each pair will enter into play: In case of real roots only the positive (O and X) ones. In the case of imaginary roots, only the ones with positive imaginary part. And finally, for the case of complex roots (there will be two complex conjugate pairs for real $\eta_z$), the ones with positive imaginary part. These two roots are going to be used in the polarizations below.
	
	\section{Polarizations}
	
	A propagating incident wave, depending on the conditions of the ambient medium is considered to be either an O- or an X-mode (in the following the labeling "O" and "X" refer to the incident field). The homogeneous system is:
	
	\begin{align*}
	\begin{pmatrix} Ac^2+K_{\perp}s^2-\eta_{0z}^2 & (K_{\perp}-A)sc-iK_{\times}c_0 & Dc+\eta_r\eta_{0z}-iK_{\times}s_0s\\
	(K_{\perp}-A)sc+iK_{\times}c_0 & As^2+K_{\perp}c^2-\eta_r^2-\eta_{0z}^2 & -Ds-iK_{\times}s_0c\\
	Dc+\eta_r\eta_{0z}+iK_{\times}s_0s & -Ds+iK_{\times}s_0c & A'-\eta_r^2
	\end{pmatrix}\begin{pmatrix}
	E_r\\ E_{\varphi}\\ E_z
	\end{pmatrix}=0
	\end{align*}
	From the homogeneous system one obtains (suitable for the O-mode in the previous case of the aligned cylinder):
	
	\begin{align*}
	&\begin{pmatrix}
	r_{Or}^P\\ r_{O\varphi}^P
	\end{pmatrix}=\begin{pmatrix}
	(K_{\perp}c_0^2+K_{\parallel}s_0^2)c^2+K_{\perp}s^2-\eta_{0z}^2 & (K_{\perp}-K_{\parallel})s_0^2sc-iK_{\times}c_0\\
	(K_{\perp}-K_{\parallel})s_0^2sc+iK_{\times}c_0 & (K_{\perp}c_0^2+K_{\parallel}s_0^2)s^2+K_{\perp}c^2-\eta^2
	\end{pmatrix}^{-1}\\
	&\begin{pmatrix}
	(K_{\perp}-K_{\parallel})s_0c_0c-\eta_r\eta_{0z}+iK_{\times}s_0s\\
	-(K_{\perp}-K_{\parallel})s_0c_0s+iK_{\times}s_0c
	\end{pmatrix}\\
	&r_{Oz}^P=1
	\end{align*}

	For the X-mode we obtain:
	
	\begin{align*}
	&r_{Xr}^P=1\\
	&\begin{pmatrix}
	r_{X\varphi}^P\\ r_{Xz}^P
	\end{pmatrix}=
	\begin{pmatrix}
	\eta^2-(K_{\perp}c_0^2+K_{\parallel}s_0^2)s^2-K_{\perp}c^2 & -(K_{\perp}-K_{\parallel})s_0c_0s+iK_{\times}s_0c\\
	-(K_{\perp}-K_{\parallel})s_0c_0s-iK_{\times}s_0c & \eta_r^2-K_{\perp}s_0^2-K_{\parallel}c_0^2
	\end{pmatrix}^{-1}\\
	&\begin{pmatrix}
	(K_{\perp}-K_{\parallel})s_0^2sc+iK_{\times}c_0\\
	\eta_r\eta_{0z}-(K_{\perp}-K_{\parallel})s_0c_0c+iK_{\times}s_0s
	\end{pmatrix}
	\end{align*}
	Note that the vectorial expression for the O-mode type of polarizations is
	\begin{align*}
	\mathbf{e}_O^P\equiv\frac{E_{Oz}}{E_0}(\hat{\mathbf{k}}_rr_{Or}^P+\hat{\mathbf{k}}_{\varphi}r_{O\varphi}^P+\hat{\mathbf{k}}_zr_{Oz}^P)\equiv e_{Oz}^P\mathbf{r}_O^P
	\end{align*}
	and respectively, for the X-mode type polarization,
	\begin{align*}
	\mathbf{e}_X^P\equiv\frac{E_{X\varphi}}{E_0}(\hat{\mathbf{k}}_rr_{Xr}^P+\hat{\mathbf{k}}_{\varphi}r_{X\varphi}^P+\hat{\mathbf{k}}_zr_{Xz}^P)\equiv e_{X\varphi}^P\mathbf{r}_X^P
	\end{align*}
	with the hat signifying unit vectors in \textbf{k}-space. It is useful to renormalize the polarization in such a way that leads in both cases to unitary complex vector of polarization, that is,
	\begin{align*}
	\mathbf{e}_O^P\equiv e_{Oz}^P\sqrt{\mathbf{r}_O^P\cdot(\mathbf{r}_O^P)^*}(\hat{\mathbf{k}}_r\hat{E}_{Or}^P+\hat{\mathbf{k}}_{\varphi}\hat{E}_{O\varphi}^P+\hat{\mathbf{k}}_z\hat{E}_{Oz}^P)\equiv e_{Oz}^P\sqrt{\mathbf{r}_O^P\cdot(\mathbf{r}_O^P)^*}\hat{\mathbf{E}}_O^P\equiv c_O^P\hat{\mathbf{E}}_O^P(\boldsymbol{\eta}_0)
	\end{align*}
	and 
	\begin{align*}
	\mathbf{e}_X^P\equiv e_{X\varphi}^P\sqrt{\mathbf{r}_X^P\cdot(\mathbf{r}_X^P)^*}(\hat{\mathbf{k}}_r\hat{E}_{X\varphi}^P+\hat{\mathbf{k}}_{\varphi}\hat{E}_{Xz}^P+\hat{\mathbf{k}}_z\hat{E}_{Xz}^P)\equiv e_{X\varphi}^P\sqrt{\mathbf{r}_X^P\cdot(\mathbf{r}_X^P)^*}\hat{\mathbf{E}}_X^P\equiv c_X^P\hat{\mathbf{E}}_X^P(\boldsymbol{\eta}_0)
	\end{align*}
	where, by definition now:
	\begin{align*}
	\hat{\mathbf{E}}_M^P\cdot(\hat{\mathbf{E}}_M^P)^*=1, \quad M=O,X
	\end{align*}
	and the corresponding normalized vectorial electric field intensities $\mathbf{e}_O^P$ and $\mathbf{e}_X^P$ are proportional to the respective unitary complex polarization vectors via the arbitrary coefficients $c_M^P$ (the polarization amplitudes) which are functions of the azimuthal angle and $\eta_{0z}$. Because of azimuthal symmetries described previously and the requirement of azimuthal continuity (azimuthal invariance under a rotation by $\pi$) they are related within pairs of modes (independently O or X). They are also periodic functions of the azimuthal angle. For this reason they can be written as a superposition of azimuthal modes which form a complete basis, that is:
	\begin{align*}
	c_{M_s}^P(\varphi_k,\eta_{kz})=\sum_{n=-\infty}^{n=\infty}\varepsilon_n^{M_s}(\eta_{kz})\text{e}^{in\varphi_k}
	\end{align*}
	where $\varepsilon_n^{M_s}(\eta_{kz})$ are going to be determined. For the incident field we always set $c_k^P=1$ for either O- or X-mode. We now calculate the following dot products:
	\begin{align*}
	\mathbf{e}_k^P\cdot\mathbf{a}_m, \quad \mathbf{e}_k^P\cdot\mathbf{b}_m, \quad \mathbf{e}_k^P\cdot\mathbf{c}_m; \quad k=O,X
	\end{align*}
	that is (from known formulas for $\mathbf{a}_m$, $\mathbf{b}_m$ and $\mathbf{c}_m$ in Cartesian form expressed in terms of the radial component and the azimuth):
	\begin{align*}
	&\mathbf{e}_k^P\cdot\mathbf{a}_m=c_k^P(\hat{E}_{kr}^P\hat{\mathbf{k}}_r+\hat{E}_{k\varphi_k}^P\hat{\boldsymbol{\varphi}}_k+\hat{E}_{kz}^P\hat{\mathbf{z}})\cdot(-\hat{\mathbf{x}}\sin\varphi_k+\hat{\mathbf{y}}\cos\varphi_k)\frac{i^{m+1}\exp(-im\varphi_k)}{\eta_{kr}}
	\end{align*}
	\begin{align*}
	\begin{split}
	\mathbf{e}_k^P\cdot\mathbf{b}_m=&-c_k^P(\hat{E}_{kr}^P\hat{\mathbf{k}}_r+\hat{E}_{k\varphi_k}^P\hat{\boldsymbol{\varphi}}_k+\hat{E}_{kz}^P\hat{\mathbf{z}})\cdot(\hat{\mathbf{x}}\eta_{kz}\cos\varphi_k+\hat{\mathbf{y}}\eta_{kz}\sin\varphi_k-\hat{\mathbf{z}}\eta_{kr})\\
	&\frac{i^{m}\exp(-im\varphi_k)}{\eta_k\eta_{kr}}
	\end{split}
	\end{align*}
	\begin{align*}
	\begin{split}
	\mathbf{e}_k^P\cdot\mathbf{c}_m=&-c_k^P(\hat{E}_{kr}^P\hat{\mathbf{k}}_r+\hat{E}_{k\varphi_k}^P\hat{\boldsymbol{\varphi}}_k+\hat{E}_{kz}^P\hat{\mathbf{z}})\cdot(\hat{\mathbf{x}}\eta_{kr}\cos\varphi_k+\hat{\mathbf{y}}\eta_{kr}\sin\varphi_k+\hat{\mathbf{z}}\eta_{kz})\\
	&\frac{i^{m+1}\exp(-im\varphi_k)}{\eta_k^2}
	\end{split}
	\end{align*}
	On the other hand, one has:
	\begin{align*}
	&\hat{\mathbf{k}}_r\cdot\hat{\mathbf{x}}=\cos\varphi_k, \quad \hat{\boldsymbol{\varphi}}_k\cdot\hat{\mathbf{x}}=-\sin\varphi_k, \quad \hat{\mathbf{z}}\cdot\hat{\mathbf{x}}=0\\
	&\hat{\mathbf{k}}_r\cdot\hat{\mathbf{y}}=\sin\varphi_k, \quad \hat{\boldsymbol{\varphi}}_k\cdot\hat{\mathbf{y}}=\cos\varphi_k, \quad \hat{\mathbf{z}}\cdot\hat{\mathbf{y}}=0\\
	&\hat{\mathbf{k}}_r\cdot\hat{\mathbf{z}}=0, \quad \hat{\boldsymbol{\varphi}}_k\cdot\hat{\mathbf{z}}=0, \quad \hat{\mathbf{z}}\cdot\hat{\mathbf{z}}=1\\
	\end{align*}
	Therefore:
	\begin{align*}
	&\mathbf{e}_k^P\cdot\mathbf{a}_m=c_k^P\hat{E}_{k\varphi_k}^P\frac{i^{m+1}\exp(-im\varphi_k)}{\eta_{kr}}\\
	&\mathbf{e}_k^P\cdot\mathbf{b}_m=c_k^P(\hat{E}_{kz}^P\eta_{kr}-\hat{E}_{kr}^P\eta_{kz})\frac{i^{m}\exp(-im\varphi_k)}{\eta_k\eta_{kr}}\\
	&\mathbf{e}_k^P\cdot\mathbf{c}_m=-c_k^P(\hat{E}_{kr}^P\eta_{kr}+\hat{E}_{kz}^P\eta_{kz})\frac{i^{m+1}\exp(-im\varphi_k)}{\eta_k^2}
	\end{align*}
	Note that for the incident wave we set the coefficients $c_k^P=1$ and replace $k$ by $"k_0"$. Note also that in all the following, the following definition holds:
	\begin{align*}
	\eta^2\equiv\eta_r^2+\eta_{0z}^2
	\end{align*}

	So, the previously obtained results are taking the form:

	\begin{align*}
	\begin{split}
	&r_{Xr}^P=1\\
	&r_{X\varphi}^P=\frac{1}{d_X}(K_{\perp}-K_{\parallel})\eta_r\eta_{0z}s_0c_0s+\left[K_{\times}^2-(K_{\perp}-K_{\parallel})(K_{\perp}-\eta_r^2)\right]css_0^2\\
	&-iK_{\times}\left[(K_{\parallel}-\eta_r^2)c_0+\eta_r\eta_{0z}s_0c\right]\\
	&r_{Xz}^P
	=\frac{1}{d_X}
	(K_{\perp}-K_{\parallel})\left[\eta_r\eta_{0z}s_0s^2+(K_{\perp}-\eta^2)c_0c\right]s_0-(K_{\perp}-\eta^2)\eta_r\eta_{0z}-K_{\times}^2s_0c_0c\\
	&-iK_{\times}(K_{\parallel}-\eta^2)s_0s\\
	& \text{where} \\
	& d_X\equiv -(K_{\perp}-\eta^2)\eta_r^2-\eta^2K_{\parallel}c_0^2+(s^2s_0^2+c_0^2)K_{\perp}K_{\parallel}\\
	&+\left\{\left[K_{\perp}(K_{\perp}-\eta_r^2)-K_{\times}^2\right]c^2-K_{\parallel}\eta_r^2s^2-K_{\perp}\eta_{0z}^2\right\}s_0^2
	\end{split}
	\end{align*}
	and
	\begin{align*}
	\begin{split}
	&r_{Or}^P=\frac{1}{d_O}(K_{\perp}-K_{\parallel})\left[\eta_r\eta_{0z}s_0s^2+(K_{\perp}-\eta^2)c_0c\right]s_0-(K_{\perp}-\eta^2)\eta_r\eta_{0z}-K_{\times}^2c_0s_0c\\
	&+iK_{\times}(K_{\parallel}-\eta^2)ss_0\\
	&r_{O\varphi}^P=\frac{1}{d_O}\left\{K_{\times}^2c_0+\left[\eta_r\eta_{0z}cs_0-(K_{\perp}-\eta_{0z}^2)c_0\right](K_{\perp}-K_{\parallel})\right\}ss_0\\
	&+iK_{\times}\left[(K_{\parallel}-\eta_{0z}^2)s_0c+\eta_r\eta_{0z}c_0\right]\\
	&r_{Oz}^P=1,\\
	& \text{where} \\
	& d_O\equiv (\eta^2-\eta_r^2s^2)(K_{\perp}-K_{\parallel})s_0^2+(K_{\perp}c_0^2+K_{\parallel}s_0^2-\eta^2)K_{\perp}-(K_{\perp}-\eta^2)\eta_{0z}^2-K_{\times}^2c_0^2
	\end{split}
	\end{align*}
	\subsection{Special case: Incidence on the plane $\mathbf{(z-z')}$}
	In this case $s=0$ and $c=1$. Thus,
	\begin{align*}
	& r_{Xr}^P=1\\
	& \begin{pmatrix}
	r_{X\varphi}^P\\ r_{Xz}^P
	\end{pmatrix}=
	\begin{pmatrix}
	\dfrac{-iK_{\times}\left[(K_{\parallel}-\eta_r^2)c_0+\eta_r\eta_{0z}s_0\right]}{(K_{\perp}-\eta^2)(K_{\perp}s_0^2+K_{\parallel}c_0^2-\eta_r^2)-K_{\times}s_0^2}\\
	\dfrac{\left[(K_{\perp}-K_{\parallel})(K_{\perp}-\eta^2)-K_{\times}^2\right]s_0c_0-(K_{\perp}-\eta^2)\eta_r\eta_{0z}}{(K_{\perp}-\eta^2)(K_{\perp}s_0^2+K_{\parallel}c_0^2-\eta_r^2)-K_{\times}s_0^2}
	\end{pmatrix}\\
	& \text{and}\\
	& \begin{pmatrix}
	r_{Or}^P\\ r_{O\varphi}^P
	\end{pmatrix}=\begin{pmatrix}
	\dfrac{\left[(K_{\perp}-K_{\parallel})c_0s_0-\eta_r\eta_{0z}\right](K_{\perp}-\eta^2)-K_{\times}^2c_0s_0}{(K_{\perp}^2-K_{\times}^2)c_0^2-\eta^2(K_{\perp}c_0^2+K_{\parallel}s_0^2)-\eta_{0z}^2(K_{\perp}-\eta^2)+K_{\parallel}K_{\perp}s_0^2}\\
	\dfrac{iK_{\times}\left[(K_{\parallel}-\eta_{0z}^2)s_0+\eta_r\eta_{0z}c_0\right]}{(K_{\perp}^2-K_{\times}^2)c_0^2-\eta^2(K_{\perp}c_0^2+K_{\parallel}s_0^2)-\eta_{0z}^2(K_{\perp}-\eta^2)+K_{\parallel}K_{\perp}s_0^2}
	\end{pmatrix}\\
	&r_{Oz}^P=1
	\end{align*}
	For the aligned cylinder these boil down to:
	\begin{align*}
	\begin{pmatrix}
	r_{Xr}^P\\ r_{X\varphi}^P\\ r_{Xz}^P
	\end{pmatrix}=
	\begin{pmatrix}
	1\\
	\dfrac{-iK_{\times}}{(K_{\perp}-\eta^2)}\\
	\dfrac{-\eta_r\eta_{0z}}{K_{\parallel}-\eta_r^2}
	\end{pmatrix}
	\text{  and  }
	\begin{pmatrix}
	r_{Or}^P\\ r_{O\varphi}^P\\ r_{Oz}^P
	\end{pmatrix}=\begin{pmatrix}
	\dfrac{-\eta_r\eta_{0z}(K_{\perp}-\eta^2)}{(K_{\perp}-\eta_{0z}^2)(K_{\perp}-\eta^2)-K_{\times}^2}\\
	\dfrac{iK_{\times}\eta_r\eta_{0z}}{(K_{\perp}-\eta_{0z}^2)(K_{\perp}-\eta^2)-K_{\times}^2}\\
	1
	\end{pmatrix}
	\end{align*}
	Note that in the aligned cylinder and for a propagation vector on the $x-z$ plane, the four roots are:
	\begin{align*}
	\begin{split}
	&\eta_r^{(\pm O)}=\pm\dfrac{1}{\sqrt{2K_{\perp}}}\left\{(K_{\perp}+K_{\parallel})(K_{\perp}-\eta_{0z}^2)-K_{\times}^2\right.\\
	&\left. +\sqrt{\left[(K_{\perp}-K_{\parallel})^2(K_{\perp}-\eta_{0z}^2)-2K_{\times}^2(K_{\perp}+K_{\parallel})\right](K_{\perp}-\eta_{0z}^2)+(4K_{\perp}K_{\parallel}+K_{\times}^2)K_{\times}^2}\right\}^{\frac{1}{2}}
	\end{split}
	\end{align*}
	\begin{align*}
	\begin{split}
	&\eta_r^{(\pm X)}=\pm\dfrac{1}{\sqrt{2K_{\perp}}}\left\{(K_{\perp}+K_{\parallel})(K_{\perp}-\eta_{0z}^2)-K_{\times}^2\right.\\
	&\left. -\sqrt{\left[(K_{\perp}-K_{\parallel})^2(K_{\perp}-\eta_{0z}^2)-2K_{\times}^2(K_{\perp}+K_{\parallel})\right](K_{\perp}-\eta_{0z}^2)+(4K_{\perp}K_{\parallel}+K_{\times}^2)K_{\times}^2}\right\}^{\frac{1}{2}}
	\end{split}
	\end{align*}
	that is, two O-type and two X-type. \textbf{Note that $\mathbf{+M (-M)}$ corresponds to $\mathbf{M_1 (M_2)}$ with $\mathbf{M=O,X}$.}
	Therefore, each pair consists of two opposite complex roots in general. From these pairs emanate the four roots in the inclined cylinder case that cease to form pairs as before but they are close to forming pairs as the inclination approaches zero. However, the four roots as functions of the azimuthal angle form "symmetry-based pairs" as we have shown before. In the following we may continue labeling the roots as O-type and X-type according to which type they fall to as the inclination goes to zero. For the field-aligned cylinder and the propagation vector on the $x-z$ plane one can easily observe that:
	\begin{align*}
	\begin{pmatrix}
	r_{X\varphi}^P(-\eta_r)\\ r_{Xz}^P(-\eta_r)
	\end{pmatrix}=
	\begin{pmatrix}
	r_{X\varphi}^P(\eta_r)\\ -r_{Xz}^P(\eta_r)
	\end{pmatrix}, \quad
	\begin{pmatrix}
	r_{Or}^P(-\eta_r)\\ r_{O\varphi}^P(-\eta_r)
	\end{pmatrix}=
	\begin{pmatrix}
	-r_{Or}^P(\eta_r)\\ -r_{O\varphi}^P(\eta_r)
	\end{pmatrix}
	\end{align*}
	In the general case, on the other hand, we have shown that
	\begin{align*}
	\eta_{kr}^{M_i}(\varphi_k)=\eta_{kr}^{M_i}(2\pi-\varphi_k)=-\eta_{kr}^{M_{3-i}}(\pi-\varphi_k)
	\end{align*}
	but nothing can be said as a general statement for the symmetries of the polarizations as functions of the azimuthal angle; unless, of course, the roots $\eta_r$ are all real (or, purely imaginary). In the following only one member of each pair will enter into play: In case of real roots only the positive (O and X ones). In the case of imaginary roots only the ones with positive imaginary part. And, finally, for the case of complex roots (there will be two complex conjugate pairs for real $\eta_z$) the ones with positive imaginary part.
	\section{Back to the field expressions}
	Normalizing the previously found general expression for the electric field, one obtains:
	\begin{align*}
	& \mathbf{e(r)}=\frac{\mathbf{E(r)}}{E_0}\\
	& =\int_{0}^{2\pi}d\varphi_k\int_{-\infty}^{\infty}d\eta_{kz}\\
	&\sum_{M=O,X}\left\{\mathbf{e}_k^M[\eta_{kr}^M(\varphi_k,\eta_{kz}),\varphi_k, \eta_{kz}]+\mathbf{e}_k^M[-\eta_{kr}^M(\varphi_k,\eta_{kz}),\varphi_k, \eta_{kz}]\right\}\text{e}^{i\eta_{kr}^M\rho\cos(\varphi-\varphi_k)+i\eta_{kz}\zeta}
	\end{align*}
	where the eigenmodes under the integral sign are pice-wise spatially constant. For the incident field on the other hand,
	\begin{align*}
	\mathbf{e}_0\mathbf{(r)}=\frac{\mathbf{E}_0\mathbf{(r)}}{E_0}=\mathbf{e}_0(\varphi_0,\eta_{0z})\text{e}^{i\eta_{0r}\rho\cos(\varphi-\varphi_0)+i\eta_{0z}\zeta}
	\end{align*}
	In terms of the cylindrical vector functions and the exponential dyadic, we have:
	\begin{align*}
	\begin{split}
	&\mathbf{e}_k^M[\eta_{kr}^M(\varphi_k,\eta_{kz}),\varphi_k, \eta_{kz}]\text{e}^{i\eta_{kr}^M\rho\cos(\varphi-\varphi_k)+i\eta_{kz}\zeta}=\\
	\sum_{m=-\infty}^{m=\infty}&
	\left\{\mathbf{e}_k^M[\eta_{kr}^M(\varphi_k,\eta_{kz}),\varphi_k, \eta_{kz}]\cdot \mathbf{a}_m[\eta_{kr}^M(\varphi_k,\eta_{kz}),\varphi_k, \eta_{kz}]\mathbf{m}_m[\rho\eta_{kr}^M(\varphi_k,\eta_{kz}),\varphi, \zeta\eta_{kz}]\right.\\
	& \left. +\mathbf{e}_k^M[\eta_{kr}^M(\varphi_k,\eta_{kz}),\varphi_k, \eta_{kz}]\cdot \mathbf{b}_m[\eta_{kr}^M(\varphi_k,\eta_{kz}),\varphi_k, \eta_{kz}]\mathbf{n}_m[\rho\eta_{kr}^M(\varphi_k,\eta_{kz}),\varphi, \zeta\eta_{kz}]\right.\\
	& \left. +\mathbf{e}_k^M[\eta_{kr}^M(\varphi_k,\eta_{kz}),\varphi_k, \eta_{kz}]\cdot \mathbf{c}_m[\eta_{kr}^M(\varphi_k,\eta_{kz}),\varphi_k, \eta_{kz}]\mathbf{l}_m[\rho\eta_{kr}^M(\varphi_k,\eta_{kz}),\varphi, \zeta\eta_{kz}]\right\}
	\end{split}
	\end{align*}
	and respectively, for the incident field:
	\begin{align*}
	\begin{split}
	&\mathbf{e}_0(\eta_{0r},\eta_{0z},\varphi_0),\text{e}^{i\eta_{0r}\rho\cos(\varphi-\varphi_0)+i\eta_{0z}\zeta}=\\
	&\sum_{m=-\infty}^{m=\infty}\left\{\mathbf{e}_0(\eta_{0r},\eta_{0z},\varphi_0)\cdot\mathbf{a}_m(\eta_{0r},\eta_{0z},\varphi_0)\mathbf{m}_m(\rho\eta_{0r},\eta_{0z}\zeta,\varphi_0)\right.\\
	& \left. +\mathbf{e}_0(\eta_{0r},\eta_{0z},\varphi_0)\cdot\mathbf{b}_m(\eta_{0r},\eta_{0z},\varphi_0)\mathbf{n}_m(\rho\eta_{0r},\eta_{0z}\zeta,\varphi_0)\right.\\
	& \left. +\mathbf{e}_0(\eta_{0r},\eta_{0z},\varphi_0)\cdot\mathbf{c}_m(\eta_{0r},\eta_{0z},\varphi_0)\mathbf{l}_m(\rho\eta_{0r},\eta_{0z}\zeta,\varphi_0)\right\} 
	\end{split}
	\end{align*}
	However,
	\begin{align*}
	\mathbf{e}_k^M[\eta_{kr}^M(\varphi_k,\eta_{kz}),\varphi_k, \eta_{kz}]&=c_M^P(\varphi_k,\eta_{kz})\hat{\mathbf{E}}_M^P[\eta_{kr}^M(\varphi_k,\eta_{kz}),\varphi_k, \eta_{kz}]\\
	& =\sum_{n=-\infty}^{n=\infty}\varepsilon_n^M(\eta_{kz})\text{e}^{in\varphi_k}\hat{\mathbf{E}}_M^P[\eta_{kr}^M(\varphi_k,\eta_{kz}),\varphi_k, \eta_{kz}]
	\end{align*}
	Note that for the aligned cylinder the polarizations and the radial component of the propagation vector do not explicitly depend on the azimuthal angle. From previous calculations we have in Cartesian form:
	\begin{align*}
	\begin{split}
	&\mathbf{a}_m=i^{m+1}\frac{\text{e}^{-im\varphi_k}}{\eta_{kr}}\begin{pmatrix}
	-\sin\varphi_k\\ \cos\varphi_k\\ 0
	\end{pmatrix}, \quad
	\mathbf{b}_m=-i^m\frac{\eta_{kz}\text{e}^{im\varphi_k}}{\eta_k\eta_{kr}}\begin{pmatrix}
	\cos\varphi_k\\ \sin\varphi_k\\ -\frac{\eta_{kr}}{\eta_{kz}}
	\end{pmatrix},\\
	&\mathbf{c}_m=-i^{m+1}\frac{\text{e}^{-im\varphi_k}}{\eta_{kr}^2}\begin{pmatrix}
	\eta_{kr}\cos\varphi_k\\ \eta_{kr}\sin\varphi_k\\ \eta_{kz}
	\end{pmatrix}
	\end{split}
	\end{align*}
	Introducing the formulas for the inner products as well as the azimuthal expansion of the polarization amplitudes, one obtains in normalized form:
	\begin{align*}
	\begin{split}
	&\mathbf{e(r)}=\sum_{m=-\infty}^{m=\infty}i^m\int_{0}^{2\pi}d\varphi_k\int_{-\infty}^{\infty}d\eta_{kz}\sum_{M=O,X}\sum_{n=-\infty}^{n=\infty}\varepsilon_n^M(\eta_{kz}\text{e}^{i(n-m)\varphi_k}\\
	&\left\{i\frac{\hat{E}_{k\varphi_k}^M}{\eta_{kr}^M}\mathbf{m}_m(\rho\eta_{kr}^M,\varphi,\zeta\eta_{kz})+\frac{\hat{E}_{kz}^M\eta_{kr}^M-\hat{E}_{kr}^M\eta_{kz}}{\eta_{kr}^M\eta_k^M}\mathbf{n}_m(\rho\eta_{kr}^M,\varphi,\zeta\eta_{kz})\right.\\
	&\left.-i\frac{\hat{E}_{kr}^M\eta_{kr}^M+\hat{E}_{kz}^M\eta_{kz}}{\left(\eta_{k}^M\right)^2}\mathbf{l}_m(\rho\eta_{kr}^M,\varphi,\zeta\eta_{kz})\right\}
	\end{split}
	\end{align*}
	For the incident,
	\begin{align*}
	\begin{split}
	&\mathbf{e}_0\mathbf{(r)}=\sum_{m=-\infty}^{m=\infty}i^m\text{e}^{-im\varphi_0}\left\{i\frac{\hat{E}_{0\varphi_0}}{\eta_{0r}}\mathbf{m}_m(\rho\eta_{0r},\varphi,\zeta\eta_{0z})+\frac{\hat{E}_{0z}\eta_{0r}-\hat{E}_{0r}\eta_{0z}}{\eta_{0r}\eta_0}\mathbf{n}_m(\rho\eta_{0r},\varphi,\zeta\eta_{0z})\right.\\
	&\left.-i\frac{\hat{E}_{0r}\eta_{0r}+\hat{E}_{0z}\eta_{0z}}{\eta_0^2}\mathbf{l}_m(\rho\eta_{0r},\varphi,\zeta\eta_{0z})\right\}
	\end{split}
	\end{align*}
	Notice that for the aligned cylinder the mode selection does not depend on the azimuthal angle and therefore, the integration over that angle will facilitate the application of orthogonality condition for the azimuthal dependence. The magnetic field can also be easily evaluated from Faraday's law:
	\begin{align*}
	\mathbf{h}\equiv \frac{\mathbf{H}}{\mathcal{H}_0}=\frac{\mathcal{E}_0}{\mathcal{H}_0}\sqrt{\frac{\varepsilon_0}{\mu_0}}\frac{1}{i}\nabla\times\mathbf{e}, \quad
	\mathbf{h}_0\equiv \frac{\mathbf{H}_0}{\mathcal{H}_0}=\frac{\mathcal{E}_0}{\mathcal{H}_0}\sqrt{\frac{\varepsilon_0}{\mu_0}}\frac{1}{i}\nabla\times\mathbf{e}_0
	\end{align*}
	Therefore:
	\begin{align*}
	& \mathbf{h(r)}=\frac{E_0}{H_0}\sqrt{\frac{\varepsilon_0}{\mu_0}}\sum_{m=-\infty}^{m=\infty}i^m\int_{0}^{2\pi}d\varphi_k\int_{-\infty}^{\infty}d\eta_{kz}\sum_{n=-\infty}^{n=\infty}\sum_{M=O,X}\varepsilon_n^M(\eta_{kz})\text{e}^{i(n-m)\varphi_k}\\
	&\left\{\frac{\hat{E}_{k\varphi_k}^{M}}{\eta_{kr}^M}\eta_k^M\mathbf{n}_m(\rho\eta_{kr}^M,\varphi,\zeta\eta_{kz})-i\frac{\hat{E}_{kz}^M\eta_{kr}^M-\hat{E}_{kr}^M\eta_{kz}}{\eta_{kr}^M}\mathbf{m}_m(\rho\eta_{kr}^M,\varphi,\zeta\eta_{kz})\right\}
	\end{align*}
	and for the incident
	\begin{align*}
	\begin{split}
	&\mathbf{h}_0\mathbf{(r)}=\frac{E_0}{H_0}\sqrt{\frac{\varepsilon_0}{\mu_0}}\sum_{m=-\infty}^{m=\infty}i^m\text{e}^{-im\varphi_0}\\
	&\left\{\frac{\hat{E}_{0\varphi_0}}{\eta_{0r}}\eta_0\mathbf{n}_m(\rho\eta_{0r},\varphi,\zeta\eta_{0z})-i\frac{\hat{E}_{0z}\eta_{0r}-\hat{E}_{0r}\eta_{0z}}{\eta_{0r}}\mathbf{m}_m(\rho\eta_{0r},\varphi,\zeta\eta_{0z})\right\}
	\end{split}
	\end{align*}
	\section{Boundary conditions}
	The vectorial ones are:
	\begin{align*}
	\hat{\mathbf{r}}\times(\mathbf{e}_{SC}+\mathbf{e}_0-\mathbf{e}_{BL})=0, \quad \hat{\mathbf{r}}\times(\mathbf{h}_{SC}+\mathbf{h}_0-\mathbf{h}_{BL})=0
	\end{align*}
	where SC and BL stand for the exterior and the interior of the blob respectively. One can easily express the exterior products involved in the boundary conditions solely in terms of the vector functions and the tangential dyadics of unit vectors: $\hat{\mathbf{z}}\hat{\mathbf{r}}$ and $\hat{\boldsymbol{\varphi}}\hat{\mathbf{z}}$:
	\begin{align*}
	& \hat{\mathbf{r}}\times\mathbf{m}_m=i(\hat{\mathbf{z}}\hat{\mathbf{r}})\cdot\mathbf{n}_m\frac{\eta_k}{\eta_{kz}}\\
	& \hat{\mathbf{r}}\times\mathbf{n}_m=i(\hat{\mathbf{z}}\hat{\mathbf{r}})\cdot\mathbf{m}_m\frac{\eta_{kz}}{\eta_k}-(\hat{\boldsymbol{\varphi}}\hat{\mathbf{z}})\cdot\mathbf{n}_m\\
	& \hat{\mathbf{r}}\times\mathbf{l}_m=(\hat{\mathbf{z}}\hat{\mathbf{r}})\cdot\mathbf{m}_m-i(\hat{\boldsymbol{\varphi}}\hat{\mathbf{z}})\cdot\mathbf{n}_m\frac{\eta_k\eta_{kz}}{\eta_{kr}^2}
	\end{align*}
	The vector functions are calculated at $a$ (the normalized radius of the cylinder) and the tangential components of the electric field are matched. By truncating the summation over $n$ and writing down the conditions for the azimuthal modes, that is for each $m$ with $m=-n_{max}$ to $m=n_{max}$, we obtain the following linear system to be solved:
	
	\begin{align*}
	&\sum_{n=-n_{max}}^{n=n_{max}}\left(a_{j,mn}^{O,BL}\varepsilon_n^{O,BL}+a_{j,mn}^{X,BL}\varepsilon_n^{X,BL}-a_{j,mn}^{O,SC}\varepsilon_n^{O,SC}-a_{j,mn}^{X,SC}\varepsilon_n^{X,SC}\right)=a_{j,m}^0, \\
	&j=1,2,3,4
	\end{align*}
	with:
	\begin{align*}
	& a_{1,mn}^{M,BL}=\int_{0}^{2\pi}\frac{d\varphi_k}{2\pi}\text{e}^{i(n-m)\varphi_k}\left(i\hat{E}_{k\varphi_k}^MJ_m^{'M}-\hat{E}_{kr}^M\frac{mJ_m^M}{\eta_{kr}^Ma}\right),\\
	& a_{1,mn}^{M,SC}=\int_{0}^{2\pi}\frac{d\varphi_k}{2\pi}\text{e}^{i(n-m)\varphi_k}\left(i\hat{E}_{k\varphi_k}^MH_m^{'(1)M}-\hat{E}_{kr}^M\frac{mH_m^{(1)M}}{\eta_{kr}^Ma}\right), \\
	&a_{1,m}^{0}=\text{e}^{-im\varphi_0}\left(i\hat{E}_{0\varphi_0}J_m^{'0}-\hat{E}_{0r}\frac{mJ_m^0}{\eta_{0r}a}\right)\\
	& a_{2,mn}^{M,BL}=\int_{0}^{2\pi}\frac{d\varphi_k}{2\pi}\text{e}^{i(n-m)\varphi_k}\left(\hat{E}_{kz}^MJ_m^{M}\right), \quad a_{2,mn}^{M,SC}=\int_{0}^{2\pi}\frac{d\varphi_k}{2\pi}\text{e}^{i(n-m)\varphi_k}\left(\hat{E}_{kz}^MH_m^{(1)M}\right),\\
	&  a_{2,m}^{0}=\text{e}^{-im\varphi_0}\left(\hat{E}_{0z}J_m^{0}\right)\\
	& a_{3,mn}^{M,BL}=\int_{0}^{2\pi}\frac{d\varphi_k}{2\pi}\text{e}^{i(n-m)\varphi_k}\left[i\left(\hat{E}_{kz}^M\eta_{kr}^M-\hat{E}_{kr}^M\eta_{0z}\right)J_m^{'M}-\hat{E}_{k\varphi_k}^M\eta_{0z}\frac{mJ_m^M}{\eta_{kr}^Ma}\right],\\
	& a_{3,mn}^{M,SC}=\int_{0}^{2\pi}\frac{d\varphi_k}{2\pi}\text{e}^{i(n-m)\varphi_k}\left[i\left(\hat{E}_{kz}^M\eta_{kr}^M-\hat{E}_{kr}^M\eta_{0z}\right)H_m^{'(1)M}-\hat{E}_{k\varphi_k}^M\eta_{0z}\frac{mH_m^{(1)M}}{\eta_{kr}^Ma}\right],\\
	& a_{3,m}^0=\text{e}^{-im\varphi_0}\left[i\left(\hat{E}_{0z}\eta_{0r}-\hat{E}_{0r}\eta_{0z}\right)J_m^{'0}-\hat{E}_{0\varphi_0}^M\eta_{0z}\frac{mJ_m^0}{\eta_{0r}a}\right],\\
	& a_{4,mn}^{M,BL}=\int_{0}^{2\pi}\frac{d\varphi_k}{2\pi}\text{e}^{i(n-m)\varphi_k}\left(\hat{E}_{k\varphi_k}^M\eta_{kr}^MJ_m^{M}\right), \\
	&a_{4,mn}^{M,SC}=\int_{0}^{2\pi}\frac{d\varphi_k}{2\pi}\text{e}^{i(n-m)\varphi_k}\left(\hat{E}_{k\varphi_k}^M\eta_{kr}^MH_m^{(1)M}\right), \quad a_{4,m}^{0}=\text{e}^{-im\varphi_0}\left(\hat{E}_{0\varphi_0}\eta_{0r}J_m^{0}\right)
	\end{align*}
	
	\section{EM fields and Poynting vector}
	\subsection{Synthesizing the fields at arbitrary $z$}
	Integrating out the $z$ dependence (denoted by the tilde over the vector functions), one obtains for the $(x-y)$ plane:
	\begin{align*}
	\begin{split}
	& \tilde{\mathbf{e}}(\rho,\varphi)_{(BL,SC)}=\sum_{m=-\infty}^{m=\infty}i^m\int_0^{2\pi}d\varphi_k\sum_{M=O,X}\sum_{n=-\infty}^{n=\infty}\varepsilon_n^{M,(BL,SC)}\text{e}^{i(n-m)\varphi_k}\\
	& \left[i\frac{\hat{E}_{k\varphi_k}^M}{\eta_{kr}^M}\tilde{\mathbf{m}}_m(\rho\eta_{kr}^M,\varphi)+\frac{\hat{E}_{kz}^M\eta_{kr}^M-\hat{E}_{kr}^M\eta_{0z}}{\eta_{kr}^M\eta_{k}^M}\tilde{\mathbf{n}}_m(\rho\eta_{kr}^M,\varphi)\right.\\
	&\left.-i\frac{\hat{E}_{kr}^M\eta_{kr}^M+\hat{E}_{kz}^M\eta_{0z}}{(\eta_k^M)^2}\tilde{\mathbf{l}}_m(\rho\eta_{kr}^M,\varphi)\right]_{(BL,SC)}\\
	& \tilde{\mathbf{h}}(\rho,\varphi)_{(BL,SC)}=\frac{E_0}{H_0}\sqrt{\frac{\varepsilon_0}{\mu_0}}\sum_{m=-\infty}^{m=\infty}i^m\int_0^{2\pi}d\varphi_k\sum_{M=O,X}\sum_{n=-\infty}^{n=\infty}\varepsilon_n^{M,(BL,SC)}\text{e}^{i(n-m)\varphi_k}\\
	& \left[\frac{\hat{E}_{k\varphi_k}^M}{\eta_{kr}^M}\eta_k^M\tilde{\mathbf{n}}_m(\rho\eta_{kr}^M,\varphi)-i\frac{\hat{E}_{kz}^M\eta_{kr}^M-\hat{E}_{kr}^M\eta_{0z}}{\eta_{kr}^M}\tilde{\mathbf{m}}_m(\rho\eta_{kr}^M,\varphi)\right]_{(BL,SC)}
	\end{split}
	\end{align*}
	Respectively for the incident field:
	\begin{align*}
	& \tilde{\mathbf{e}}_0(\rho,\varphi)=\sum_{m=-\infty}^{m=\infty}i^m\text{e}^{-im\varphi_0}\\
	& \left[i\frac{\hat{E}_{0\varphi_0}}{\eta_{0r}}\eta_k^M\tilde{\mathbf{m}}_m(\rho\eta_{0r},\varphi)+\frac{\hat{E}_{0z}\eta_{0r}-\hat{E}_{0r}\eta_{0z}}{\eta_{0r}\eta_{0}}\tilde{\mathbf{n}}_m(\rho\eta_{0r},\varphi)-i\frac{\hat{E}_{0r}\eta_{0r}+\hat{E}_{0z}\eta_{0z}}{\eta_0^2}\tilde{\mathbf{l}}_m(\rho\eta_{0r},\varphi)\right]\\
	& \tilde{\mathbf{h}}_0(\rho,\varphi)=\frac{E_0}{H_0}\sqrt{\frac{\varepsilon_0}{\mu_0}}\sum_{m=-\infty}^{m=\infty}i^m\text{e}^{-im\varphi_0}\\
	&\left[\frac{\hat{E}_{0\varphi_0}}{\eta_{0r}}\eta_0\tilde{\mathbf{n}}_m(\rho\eta_{0r},\varphi)-i\frac{\hat{E}_{0z}\eta_{0r}-\hat{E}_{0r}\eta_{0z}}{\eta_{0r}}\tilde{\mathbf{m}}_m(\rho\eta_{0r},\varphi)\right]
	\end{align*}
	By introducing the respective expressions for the vector cylinder functions, one obtains:
	\begin{align*}
	& \tilde{\mathbf{e}}(\rho,\varphi)_{(BL,SC)}=\sum_{m=-\infty}^{m=\infty}i^m\text{e}^{im\varphi}\int_0^{2\pi}d\varphi_k\sum_{M=O,X}\sum_{n=-\infty}^{n=\infty}\varepsilon_n^{M,(BL,SC)}\text{e}^{i(n-m)\varphi_k}\\
	& \left[-\left(\hat{E}_{k\varphi_k}^M\frac{mZ_m^M}{\eta_{kr}^M\rho}+i\hat{E}_{kr}^MZ_m^{'M}\right)\hat{\mathbf{r}}+\left(\hat{E}_{kr}^M\frac{mZ_m^M}{\eta_{kr}^M\rho}-i\hat{E}_{k\varphi_k}^MZ_m^{'M}\right)\hat{\boldsymbol{\varphi}}+\hat{E}_{kz}^MZ_m^M\hat{\mathbf{z}}\right]_{(BL,SC)}\\
	\begin{split}
	& \tilde{\mathbf{h}}(\rho,\varphi)_{(BL,SC)}=\frac{E_0}{H_0}\sqrt{\frac{\varepsilon_0}{\mu_0}}\sum_{m=-\infty}^{m=\infty}i^m\text{e}^{im\varphi}\int_0^{2\pi}d\varphi_k\sum_{M=O,X}\sum_{n=-\infty}^{n=\infty}\varepsilon_n^{M,(BL,SC)}\text{e}^{i(n-m)\varphi_k}\\
	& \left\{\left[i\hat{E}_{k\varphi_k}^M\eta_{0z}Z_m^{'M}+\left(\hat{E}_{kz}^M\eta_{kr}^M-\hat{E}_{kr}^M\eta_{0z}\right)\frac{mZ_m^M}{\eta_{kr}^M\rho}\right]\hat{\mathbf{r}}\right.\\
	&\left.+\left[i\left(\hat{E}_{kz}^M\eta_{kr}^M-\hat{E}_{kr}^M\eta_{0z}\right)Z_m^{'M}-\hat{E}_{k\varphi_k}^M\eta_{0z}\frac{mZ_m^M}{\eta_{kr}^M\rho}\right]\hat{\boldsymbol{\varphi}}+\hat{E}_{k\varphi_k}^M\eta_{kr}^MZ_m^M\hat{\mathbf{z}}\right\}_{(BL,SC)}
	\end{split}
	\end{align*}
	Respectively for the incident field:
	\begin{align*}
	& \tilde{\mathbf{e}}_0(\rho,\varphi)=\sum_{m=-\infty}^{m=\infty}i^m\text{e}^{im(\varphi-\varphi_0)}\\
	&\left[-\left(\hat{E}_{0\varphi_0}\frac{mJ_m^0}{\eta_{0r}\rho}+i\hat{E}_{0r}J_m^{'0}\right)\hat{\mathbf{r}}+\left(\hat{E}_{0r}\frac{mJ_m^0}{\eta_{0r}\rho}-i\hat{E}_{0\varphi_0}J_m^{'0}\right)\hat{\boldsymbol{\varphi}}+\hat{E}_{0z}J_m^0\hat{\mathbf{z}}\right]\\
	\begin{split}
	& \tilde{\mathbf{h}}_0(\rho,\varphi)=\frac{E_0}{H_0}\sqrt{\frac{\varepsilon_0}{\mu_0}}\sum_{m=-\infty}^{m=\infty}i^m\text{e}^{im(\varphi-\varphi_0)}
	\left\{\left[i\hat{E}_{0\varphi_0}\eta_{0z}J_m^{'0}+\left(\hat{E}_{0z}\eta_{0r}-\hat{E}_{0r}\eta_{0z}\right)\frac{mJ_m^0}{\eta_{0r}\rho}\right]\hat{\mathbf{r}}\right.\\
	& \left.+\left[i\left(\hat{E}_{0z}\eta_{0r}-\hat{E}_{0r}\eta_{0z}\right)J_m^{'0}-\hat{E}_{0\varphi_0}\eta_{0z}\frac{mJ_m^0}{\eta_{0r}\rho}\right]\hat{\boldsymbol{\varphi}}+\hat{E}_{0\varphi_0}\eta_{0r}J_m^0\hat{\mathbf{z}}\right\}
	\end{split}
	\end{align*}
	\subsection{Calculating the time-independent Poynting vector}
	Now that all the electric and magnetic field components for every mode and all regions (incident fields, scattered fields and blob fields), the Poynting vector can be easily calculated by using the well-known formula: 
	
	\begin{align*}
	\tilde{\mathbf{s}}=\dfrac{1}{2}\text{Re}\{\tilde{\mathbf{e}} \times \tilde{\mathbf{h}}^*\}
	\end{align*}
	
	The last equation is used to export all Poynting vector components for all waves and of course, is the one from which the numerical results are exported and plotted (in the next section). 
	
\section{Numerical results}
	
	By using the above formulas, one can achieve results for a variety of different scattering processes. In some cases, it is necessary to change the value of one only component while the others are kept the same, in order to easily understand the effect of each one of them on the radio frequency scattering.

	\subsection{The way the blob radius affects RF scattering}

		Starting with the blob radius, one way to study the effect of the length of the cylinder radius on the scattering process, is by plotting figures of the time-independent Poynting vector, especially of the component which points to the forward direction. Both presented (in the figure 3) cases refer to an ambient density of $10^{19}$ $m^{-3}$ (left) and a blob density of $5.0 \times 10^{19}$ $m^{-3}$ (right), magnetic field inclination zero and azimuth of the incident wave in the cylinder coordinate system zero. It must be noted that the frequency of the RF wave has a major role in the scattering effects. The effects are different when the wavelength is of bigger, smaller or about the same size compared to the fluctuation. Figure 3 shows the Electron Cyclotron case, at $170$ $GHz$, which has a wavelength of about $1.7$ $mm$ (smaller enough than the blob's dimensions). On the other hand, for the Low Hybrid waves at $4.6$ $GHz$, the wavelength is much larger - about $6.5$ $cm$, which is much bigger than the fluctuation's dimensions, too.\\ 
		
		\begin{figure}
			\centerline{\includegraphics[scale=0.55]{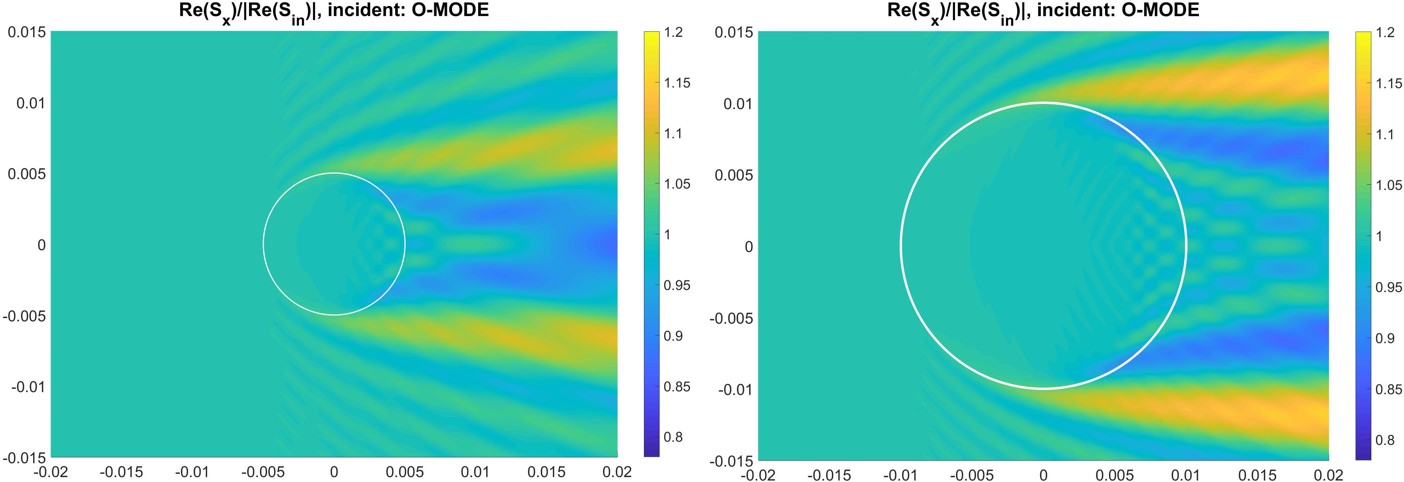}}
			\caption{Poynting flux in the forward direction, frequency $170$ $GHz$, blob radius $5$ $mm$ (left) and $10$ $mm$ (right), ambient density $10^{19}$ $m^{-3}$ and blob density $1.5 \times 10^{19}$ $m^{-3}$, magnetic field inclination $0^o$, azimuth $0^o$}
			\label{fig:ka}
		\end{figure}


	\subsection{The way the blob density contrast affects RF scattering}

		Another interesting parameter for study, is the density contrast between the cylindrical filament and the background environment. The electrons density of the filament $n_{fila}$ can be much different compared to the ambient electrons density $n_{ambi}$, so that the relative density contrast between the blob and the ambient electrons density can practically vary enough. A typical experimental range of values for $\delta n\equiv |n_{fila}-n_{ambi}|/n_{fila}$ is inside $(0.05,1)$. In figure 4, there are two cases for the Electron Cyclotron waves at frequency $170$ $GHz$ for the same cylindrical blob radius, equal to $10$ $mm$. There is neither magnetic field inclination with respect to the cylinder axis, nor azimuth taken into account. The left side is for ambient density $10^{19}$ $m^{-3}$ and blob density $5.0 \times 10^{19}$ $m^{-3}$, while the right side is for the same ambient density, but blob density is higher equal to $15 \times 10^{19}$ $m^{-3}$, so that the density contrast is different.
		
		\begin{figure}
		\centerline{\includegraphics[scale=0.55]{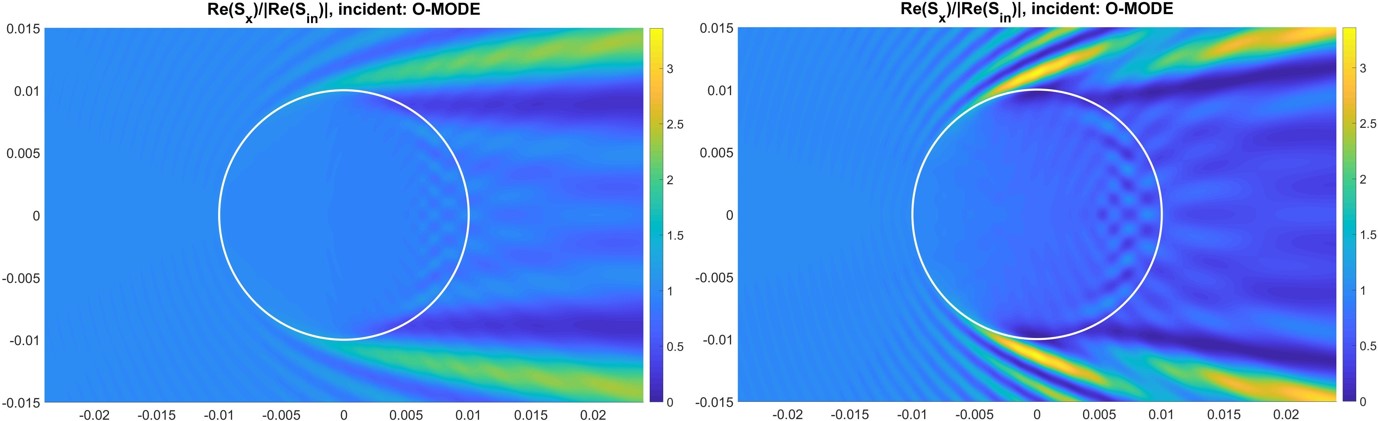}}
		\caption{Poynting flux in the forward direction, frequency 170 $GHz$, blob radius 10 $mm$, ambient density $10^{19}$ $m^{-3}$ and blob density $5.0 \times 10^{19}$ $m^{-3}$ (left) and $15 \times 10^{19}$ $m^{-3}$ (right), magnetic field inclination $0^o$, azimuth $0^o$}
		\label{fig:ka}
		\end{figure}

	\subsection{RF scattering with magnetic field inclination with respect to the cylinder axis}
		
	In figure 5, one can see the effect that the magnetic field inclination with respect to the cylinder axis has on the radiofrequency waves scattering. The magnetic field is chosen to be in angle of $5^o$ with respect to the cylinder axis, while the chosen frequency of $170$ $GHz$ is referring again to the Electron Cyclotron waves case, the blob radius is equal to $10$ $mm$, the ambient density is $2.0 \times 10^{20}$ $m^{-3}$ the blob density is $3.0 \times 10^{20}$ $m^{-3}$ and there is no azimuth taken into account. 
		\begin{figure}
		\centerline{\includegraphics[scale=0.55]{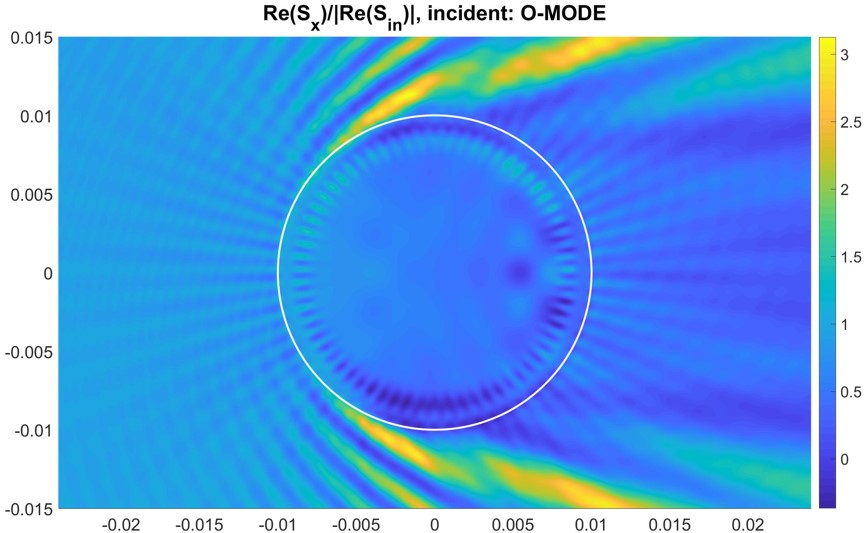}}
		\caption{Poynting flux in the forward direction, frequency 170 $GHz$, blob radius 10 $mm$, ambient density $2.0 \times 10^{20}$ $m^{-3}$ and blob density $3.0 \times 10^{20}$ $m^{-3}$, magnetic field inclination $5^o$, azimuth $0^o$}
		\label{fig:ka}
		\end{figure}

	\subsection{The way the magnetic field inclination affects RF scattering for non-zero azimuth}
	
	The effect that the magnetic field inclination has on RF scattering, can also be studied in parallel with non-zero azimuth angle. So, in figure 6 the frequency is at $170$ $GHz$ (EC), the blob radius is equal to $10$ $mm$, the ambient density is $10^{19}$ $m^{-3}$ the blob density is $1.5 \times 10^{19}$ $m^{-3}$, the azimuth is $30^o$ and the magnetic field inclination is $0^o$ (left picture) and $30^o$ (right picture).

		\begin{figure}
		\centerline{\includegraphics[scale=0.55]{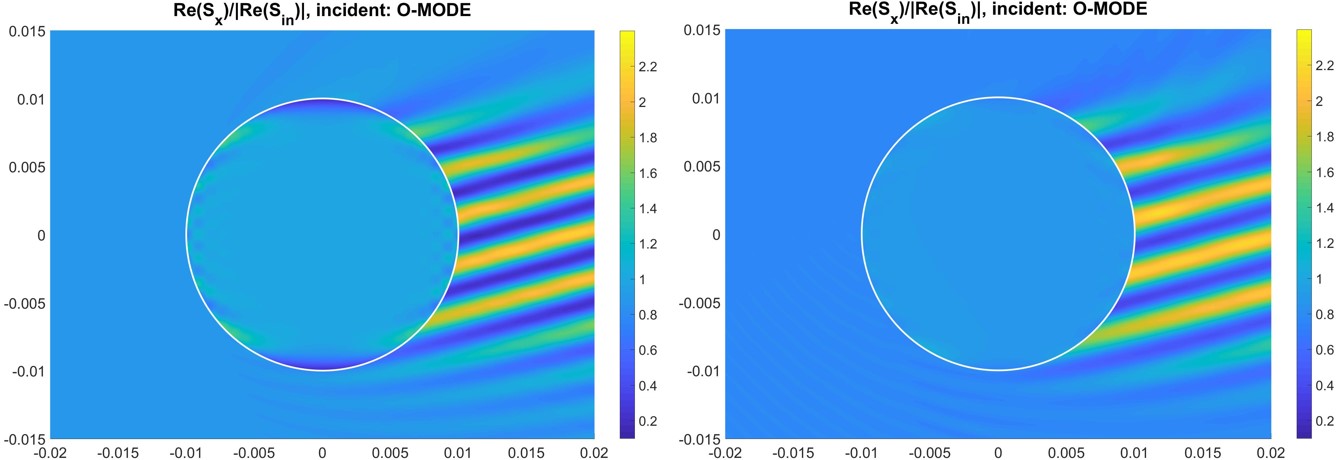}}
		\caption{Poynting flux in the forward direction, frequency 170 $GHz$, blob radius 10 $mm$, ambient density $10^{19}$ $m^{-3}$ and blob density $1.5 \times 10^{19}$ $m^{-3}$, magnetic field inclination $0^o$ (left) and $30^o$ (right), azimuth $30^o$}
		\label{fig:ka}
		\end{figure}

\goodbreak

	
\bibliographystyle{jpp}
	
\bibliography{jpp-instructions}
	
\end{document}